\documentclass[aps,pra,superscriptaddress,amssymb,onecolumn,notitlepage]{revtex4-1}
\pdfoutput=1
\usepackage{amsmath}    
\usepackage{graphicx}   
\usepackage{verbatim}   
\usepackage{color}      
\usepackage{subfigure}  
\usepackage{hyperref}   
\usepackage{booktabs}	
\usepackage{soul}
\usepackage{todonotes}
\usepackage[textwidth=7.1in,top=1.25in]{geometry}

\usepackage{amsthm} 
\usepackage{dsfont} 

\newcommand{\ket}[1]{\left|#1\right\rangle}

\newcommand{\be}{\begin{equation}}
\newcommand{\ee}{\end{equation}}

\definecolor{darkgreen}{rgb}{0.0, 0.5, 0.13}

\definecolor{tekito}{rgb}{0.9, 0.1, 0.2}

\newcommand{\fref}[1]{Supplementary Fig.~\ref{fig:#1}}
\newcommand{\eref}[1]{Supplementary Eq.~(\ref{eq:#1})}

\newcommand{\beginsupplement}{
	\setcounter{table}{0}
	\renewcommand{\thetable}{S\arabic{table}}
	\setcounter{figure}{0}
	\renewcommand{\thefigure}{S\arabic{figure}}
	\setcounter{equation}{0}
	\renewcommand{\theequation}{S\arabic{equation}}
}
\begin{document}

\title{Quantum Zeno subspaces by repeated multi-spin projections}
\author{N. Kalb}
\author{J. Cramer}

\affiliation{QuTech, Delft University of Technology, P.O. Box 5046, 2600 GA Delft, The Netherlands}
\affiliation{Kavli Institute of Nanoscience Delft, Delft University of Technology, P.O. Box 5046, 2600 GA Delft, The Netherlands}

\author{M. Markham}
\author{D.J. Twitchen}
\affiliation{Element Six Innovation, Fermi Avenue, Harwell Oxford, Didcot, Oxfordshire OX11 0QR, United Kingdom}

\author{R. Hanson}

\author{T. H. Taminiau}

\affiliation{QuTech, Delft University of Technology, P.O. Box 5046, 2600 GA Delft, The Netherlands}
\affiliation{Kavli Institute of Nanoscience Delft, Delft University of Technology, P.O. Box 5046, 2600 GA Delft, The Netherlands}

\begin{abstract}
	Repeated observations inhibit the coherent evolution of quantum states through the quantum Zeno effect. In multi-qubit systems this effect provides new opportunities to control complex quantum states. Here, we experimentally demonstrate that repeatedly projecting joint observables of multiple spins creates coherent quantum Zeno subspaces and simultaneously suppresses dephasing caused by the environment. We encode up to two logical qubits in these subspaces and show that the enhancement of the dephasing time with increasing number of projections follows a scaling law that is independent of the number of spins involved. These results provide new insights into the interplay between frequent multi-spin measurements and non-Markovian noise and pave the way for tailoring the dynamics of multi-qubit systems through repeated projections.
\end{abstract}

\maketitle

The quantum Zeno effect restricts the evolution of repeatedly observed quantum systems. For a two-dimensional system the state simply is frozen in one of two eigenstates of the measurement operator \cite{misra_zenos_1977,itano_quantum_1990,wolters_quantum_2013,kakuyanagi_observation_2015,peise_interactionfree_2015,slichter_quantum_2015,patil_measurementinduced_2015,wilkinson_experimental_1997,fischer_observation_2001}. In multi-dimensional systems, however, Zeno subspaces are formed that can contain complex quantum states and dynamics: repeated observations create a barrier that blocks coherent evolution between subspaces, but leaves coherences and dynamics within those subspaces intact \cite{facchi_quantum_2002}. Analogous effects can also be realized through coherent control pulses or strong driving fields that decouple transitions between the subspaces \cite{facchi_unification_2004,viola_dynamical_1998,dhar_preserving_2006,zheng_experimental_2013,singh_experimental_2014,zhong_optically_2015,maurer_roomtemperature_2012,lange_universal_2010} . Pioneering experiments have highlighted that the resulting non-trivial dynamics can be used to prepare exotic quantum states \cite{leghtas_confining_2015,bretheau_quantum_2015,signoles_confined_2014,barontini_deterministic_2015,schafer_experimental_2014}. However, the opportunities to tailor the dynamics of multi-qubit systems by restricting coherent evolution have remained unexplored.


Here we show that repeated multi-spin projections on individually controlled spins create quantum Zeno subspaces that can encode multiple logical qubits while suppressing dephasing caused by the environment. We realize these repeated projections for up to three nuclear spins in diamond using the optical transition of a nearby electron spin. We then encode up to two logical qubits - including entangled states of logical qubits - and show that increasing the frequency of the projections supresses the dephasing of quantum states. Finally, we theoretically derive and experimentally verify a scaling law that shows that the increase in dephasing time is independent of the number of spins involved. 


Our system consists of three $^\mathrm{13}$C spins ($I = \frac{1}{2}$) surrounding a single nitrogen vacancy (NV) centre ($\ket{0}_\mathrm{NV} : m_\mathrm{s}=0$ and $\ket{1}_\mathrm{NV} : m_\mathrm{s}=-1$) in diamond (see Supplementary Note 1).  The natural evolution of the $^\mathrm{13}$C spins is dominated by dephasing due to the slowly fluctuating surrounding bath of $^\mathrm{13}$C spins (dephasing times $T_2^* = 12.4(9)$, $8.2(7)$ and $21(1)\,\mathrm{ms}$  for spin 1, 2, and 3 respectively) \cite{cramer_repeated_2015}. Because the fluctuations are quasi-static, the Hamiltonian in a given experiment is $H=\sum_{i=1}^{k}\Delta_i\hat{\sigma}_{\mathrm{z},i}$, with $k$ the number of spins and the detuning $\Delta_i$ for spin $i$ drawn from a Gaussian distribution of width $\sqrt{2}/T_2^*$. We denote the Pauli operators as  $\hat{\sigma}_\mathrm{x}$, $\hat{\sigma}_\mathrm{y}$, $\hat{\sigma}_\mathrm{z}$ and the identity as  $\hat{I}$.


The quantum Zeno effect arises when an observable $\hat{O}$ is projected (superoperator $M(\hat{O})$), so that the system’s density matrix ($\rho_\mathrm{s}$) is left in block-diagonal form with respect to the projectors $P_\pm = (\hat{I}\pm \hat{O})/2$ \cite{facchi_quantum_2002}:

\be
M(\hat{O})\rho_\mathrm{s} = P_+\rho_\mathrm{s}P_+ + P_-\rho_\mathrm{s}P_- = \frac{\rho_\mathrm{s}+\hat{O}\rho_\mathrm{s}\hat{O}}{2}.
\ee
Repeatedly projecting observable $\hat{O}$ thus inhibits coherent evolution between the two eigenspaces of $\hat{O}$. We choose joint multi-spin observables of the form $\hat{O}=\hat{\sigma}_\mathrm{x}^{\otimes k}$, which anti-commute with all terms in the Hamiltonian $H$, so that rapid projections ideally result in the effective Zeno Hamiltonian $H_\mathrm{Zeno} = P_\pm H P_\pm = 0$ \cite{facchi_quantum_2002}. Applying these projections therefore suppresses dephasing for each nuclear spin, but leaves quantum states and driven dynamics inside the two subspaces untouched (Fig. 1a). 


To investigate quantum Zeno subspaces we use the following experimental sequence (Fig. 1b). We first initialize the nuclear spins in the desired state and prepare the electron spin in $\ket{1}_\mathrm{NV}$. Leaving the electron in $\ket{1}_\mathrm{NV}$ creates a different frequency shift for each $^\mathrm{13}$C spin that suppresses resonant flip-flop interactions during idle time \cite{bloembergen_interaction_1949}. We then apply a total of $N$ projections that are equally distributed in time. Finally the nuclear spin state is read out using the electron spin as an ancilla \cite{robledo_highfidelity_2011,taminiau_universal_2014,waldherr_quantum_2014,dreau_single-shot_2013,jiang_repetitive_2009,neumann_single-shot_2010}. Here we consider the case of an even number of projections $N$, while the results for odd $N$ are discussed in Supplementary Fig. 1. The total evolution time $\tau$ is defined from the end of the initialization to the start of the read-out. We subtract the time that control operations are applied to the nuclear spins (averaged over all spins), as dephasing might be suppressed during driving.


We experimentally realize repeated multi-spin projections on the $^\mathrm{13}$C spins by using the NV electron spin as an ancilla spin (Fig. 1c). First, we entangle the NV electron spin state with the projections on the eigenspaces of $\hat{O}$ ($\langle\hat{O}\rangle = +1$ or $-1$), so that the combined state is $\alpha \ket{\langle \hat{O} \rangle = +1}\ket{0}_\mathrm{NV} + \beta \ket{\langle \hat{O} \rangle = -1}\ket{1}_\mathrm{NV}$ \cite{cramer_repeated_2015,pfaff_demonstration_2013}. Second, we apply an optical excitation that is resonant only if the electron-spin state is $\ket{1}_\mathrm{NV}$ ('reset') \cite{robledo_highfidelity_2011}, which projects the quantum state and re-initializes the NV electron spin in $\ket{0}_\mathrm{NV}$ through optical pumping (Fig. 1d). Note that it is not required to extract or record the outcome of the optical measurement. To mitigate extra dephasing caused by the stochastic nature of the optical re-initialization (time constant of $\sim 1\,\mathrm{\mu s}$), we use $^\mathrm{13}$C spins with a NV-$^\mathrm{13}$C hyperfine coupling that is small compared to the inverse of the time constant for re-initialization (all couplings are below $2\pi \cdot 50\,\mathrm{kHz}$) \cite{blok_quantum_2015}. In addition, we design the gate sequence so that $\ket{0}_\mathrm{NV}$ is associated with the subspace of the initial nuclear state: ideally the electron spin is never optically excited and the projection constitutes a null measurement. 


To illustrate the quantum Zeno effect and to benchmark our system, we first consider a single $^\mathrm{13}$C spin and study the dephasing of the superposition state $\ket{X} \equiv (\ket{0}+\ket{1})/\sqrt{2}$ for $\hat{O} = \hat{\sigma}_\mathrm{x}$ (Fig. 2a). We initialize the $^\mathrm{13}$C spin in $\ket{X}$ with an initial state fidelity of $0.95(2)$ and apply up to $N=16$ projections. For a fixed total evolution time of $40\,\mathrm{ms}$ , we observe a significant increase of the state fidelity with an increasing number of projections (Fig. 2b). The complete time traces show that the dephasing time increases as more projections are applied (Fig. 2c); the superposition state is protected by the quantum Zeno effect. In this example, however, the Zeno subspaces contain just a single state and therefore cannot encode general quantum states.


We next investigate Zeno subspaces that can contain an arbitrary two-dimensional quantum state, i.e. a complete logical quantum bit, by performing joint projections on two $^\mathrm{13}$C spins. We set the joint observable $\hat{O} = \hat{\sigma}_\mathrm{x}\hat{\sigma}_\mathrm{x}$, so that the four-dimensional state space is divided into two coherent two-level subspaces (Fig. 3a). In these subspaces a logical qubit that can hold an arbitrary quantum state can be defined as $\ket{\psi}_\mathrm{L} = \alpha\ket{0}_\mathrm{L} + \beta\ket{1}_\mathrm{L}$, with $\ket{1}_\mathrm{L}= \ket{X,X}$ and $\ket{1}_\mathrm{L}=\ket{-X,-X}$, and with logical operators $\hat{Z}_\mathrm{L} = \hat{\sigma}_\mathrm{x} \hat{I}$ and $\hat{X}_\mathrm{L} = \hat{\sigma}_\mathrm{z} \hat{\sigma}_\mathrm{z} $. Note that logical-qubit superposition states are generally entangled states of the two $^\mathrm{13}$C spins.


We characterize the storage of quantum states by preparing all six logical basis states {$\ket{0}_\mathrm{L}$,$\ket{1}_\mathrm{L}$,$(\ket{0}_\mathrm{L}\pm \ket{1}_\mathrm{L})/\sqrt{2}$,$(\ket{0}_\mathrm{L}\pm i \ket{1}_\mathrm{L})/\sqrt{2}$} and averaging the final logical state fidelities (Fig. 3b). The logical qubit without projections shows the same decay as a single $^\mathrm{13}$C spin, but with a slightly reduced initial fidelity ($F=0.89(1)$) due to the overhead of creating the entangled states $\ket{\psi}_\mathrm{L}$. Applying projections of the joint-observable $\hat{\sigma}_\mathrm{x} \hat{\sigma}_\mathrm{x}$ strongly suppresses the dephasing by the environment, while preserving the logical qubit states. As a result, the average state fidelity for the logical qubit surpasses the best $^\mathrm{13}$C nuclear spin used, while still remaining above the threshold of $2/3$ for the storage of quantum states \cite{massar_optimal_1995}. This result demonstrates the suppression of the dephasing of a complete logical qubit through the quantum Zeno effect.


Interestingly, preserving the logical qubit does not actually require the coherence of the second spin to be maintained, as follows from the logical operator $\hat{Z}_\mathrm{L} = \hat{\sigma}_\mathrm{x} \hat{I}$. To show that the complete two-spin state is preserved, including entanglement between the two nuclear spins, we measure the average state fidelity with the ideal two-spin state for the four entangled initial states as a function of time (Fig. 3c). The duration for which genuine entanglement persists (two-spin state fidelity $> 0.5$) is extended for $N=2$, $4$ and $6$ projections compared to the case without any projections, indicating that the barrier introduced by the projections inhibits dephasing for any two-spin state within the Zeno subspace.


Realizing Zeno subspaces with even more dimensions enables the exploration of complex states of multiple logical qubits within the subspaces. We include a third nuclear spin and set $\hat{O} = \hat{\sigma}_\mathrm{x} \hat{\sigma}_\mathrm{x} \hat{\sigma}_\mathrm{x}$ to create a protected four-dimensional subspace, which can host two logical qubits defined by the logical operators $\hat{Z}_\mathrm{L1} = \hat{\sigma}_\mathrm{x} \hat{I} \hat{\sigma}_\mathrm{x}$, $\hat{X}_\mathrm{L1} = \hat{I} \hat{\sigma}_\mathrm{z} \hat{\sigma}_\mathrm{z}$ and $\hat{Z}_\mathrm{L2} = \hat{I} \hat{\sigma}_\mathrm{x} \hat{\sigma}_\mathrm{x}$, $\hat{X}_\mathrm{L2} = \hat{\sigma}_\mathrm{z} \hat{I} \hat{\sigma}_\mathrm{z}$ (Fig. 4a). Each pure state within the $\langle \hat{O} \rangle = +1$ subspace can be expressed in terms of the logical two-qubit states:

\be
\alpha \ket{X,X,X} + \beta \ket{-X,X,-X}+ \gamma \ket{X,-X,-X}+ \delta \ket{-X,-X,X} =
\alpha \ket{0,0}_\mathrm{L} + \beta \ket{0,1}_\mathrm{L}+ \gamma \ket{1,0}_\mathrm{L}+ \delta \ket{1,1}_\mathrm{L}.
\ee


To investigate the inhibition of dephasing of the two logical qubits by repeated projections we prepare three different logical states: the logical eigenstate state $\ket{0,0}_\mathrm{L}$, the logical superposition state $\ket{X,0}_\mathrm{L} = (\ket{0,0}_\mathrm{L} + \ket{1,0}_\mathrm{L})/\sqrt{2}$ and the entangled logical state $\ket{\Phi^+}_\mathrm{L} = (\ket{0,0}_\mathrm{L} + \ket{1,1}_\mathrm{L})/\sqrt{2}$. Preserving this set of states requires repeated projections of the three-spin operator $\hat{\sigma}_\mathrm{x} \hat{\sigma}_\mathrm{x} \hat{\sigma}_\mathrm{x}$ since they are not eigenstates of a single two-spin operator. 


The logical state fidelities for all three states show a clear prolongation of the decay times for $N=2$ and $N=4$ three-spin projections (Fig. 4b). Moreover, for a range of evolution times, the absolute logical state fidelities are increased despite the initial loss of fidelity due to the complexity of the experimental sequence ($33$ two-qubit gates for $N=4$, which in total require $1276$ refocusing pulses on the electron spin). These results confirm that the introduced three-spin projections inhibit dephasing of the individual spins while preserving the two logical qubits in a quantum Zeno subspace. 

To gain a detailed quantitative understanding of the quantum-Zeno effect for multi-spin projections, we derive a complete analytical description for the evolution. We model the projections as instantaneous and the noise as a quasi-static Gaussian (i.e. non-Markovian) frequency detuning, independent for each nuclear spin. We find an analytic solution for the decay of the expectation value of observables that are sensitive to dephasing (for $N$ projections and total evolution time $\tau$): 
\be
\frac{A}{2^{N+1}}\sum\limits_{l=0}^{N+1}\binom{N+1}{l}e^{-\left( \frac{t_{Nl}}{T^*_{2,\mathrm{eff}}}\right)^2} \quad \mathrm{with} \quad t_{Nl} = \tau - \frac{2l}{N+1}\tau.
\ee
Here $A\leq 1$ is the initial amplitude determined by experimental fidelities and $1/T^*_{2,\mathrm{eff}} = \sqrt{\sum_{i=1}^k (1/T^*_{2,i})^2}$ is an effective joint decay rate of all involved spins. This result is valid for any system size, i.e. number of spins, and number of projections $N$. A detailed derivation of Eq. (3) is given in Supplementary Note 2. 

We fit all experimental data in Figs. 2-4 (for $N=$ even) and Supplementary Fig. 1 (for $N=$ odd) to Eq. (3) with $A$, $T_{2,\mathrm{eff}}^*$ and an offset, to account for the fact that two out of six cardinal states are insensitive to dephasing, as free parameters. We find good agreement with the experimentally obtained dephasing curves (see Supplementary Table 1 for all fit values). To analyze the increase of the decay time with increasing number of projections we compile the extracted values from all experiments with 1, 2 and 3 nuclear spins in Fig. 5. The results reveal a scaling law that is independent of the number of spins involved, in good quantitative agreement with our theoretical model.


In conclusion, we have demonstrated that repeatedly projecting joint-observables of multi-spin systems creates quantum Zeno subspaces that can hold complex quantum states, and that these Zeno subspaces are resilient to environmental dephasing. Our results give direct insight in the physics of repeated multi-spin measurements under non-Markovian noise environments. They are also of practical relevance in the context of quantum error correction and detection codes, in which errors are detected through repeated measurements of joint observables \cite{cramer_repeated_2015,terhal_quantum_2015}. Moreover, the demonstrated methods pave the way for investigating the effect of repeated measurements in various noise environments and for exploring and engineering complex dynamics of multi-spin systems under tailored decoherence \cite{barreiro_opensystem_2011}.

\bibliography{arXiv}

\begin{thebibliography}{37}%
\makeatletter
\providecommand \@ifxundefined [1]{%
 \@ifx{#1\undefined}
}%
\providecommand \@ifnum [1]{%
 \ifnum #1\expandafter \@firstoftwo
 \else \expandafter \@secondoftwo
 \fi
}%
\providecommand \@ifx [1]{%
 \ifx #1\expandafter \@firstoftwo
 \else \expandafter \@secondoftwo
 \fi
}%
\providecommand \natexlab [1]{#1}%
\providecommand \enquote  [1]{``#1''}%
\providecommand \bibnamefont  [1]{#1}%
\providecommand \bibfnamefont [1]{#1}%
\providecommand \citenamefont [1]{#1}%
\providecommand \href@noop [0]{\@secondoftwo}%
\providecommand \href [0]{\begingroup \@sanitize@url \@href}%
\providecommand \@href[1]{\@@startlink{#1}\@@href}%
\providecommand \@@href[1]{\endgroup#1\@@endlink}%
\providecommand \@sanitize@url [0]{\catcode `\\12\catcode `\$12\catcode
  `\&12\catcode `\#12\catcode `\^12\catcode `\_12\catcode `\%12\relax}%
\providecommand \@@startlink[1]{}%
\providecommand \@@endlink[0]{}%
\providecommand \url  [0]{\begingroup\@sanitize@url \@url }%
\providecommand \@url [1]{\endgroup\@href {#1}{\urlprefix }}%
\providecommand \urlprefix  [0]{URL }%
\providecommand \Eprint [0]{\href }%
\providecommand \doibase [0]{http://dx.doi.org/}%
\providecommand \selectlanguage [0]{\@gobble}%
\providecommand \bibinfo  [0]{\@secondoftwo}%
\providecommand \bibfield  [0]{\@secondoftwo}%
\providecommand \translation [1]{[#1]}%
\providecommand \BibitemOpen [0]{}%
\providecommand \bibitemStop [0]{}%
\providecommand \bibitemNoStop [0]{.\EOS\space}%
\providecommand \EOS [0]{\spacefactor3000\relax}%
\providecommand \BibitemShut  [1]{\csname bibitem#1\endcsname}%
\let\auto@bib@innerbib\@empty
\bibitem [{\citenamefont {Misra}\ and\ \citenamefont
  {Sudarshan}(1977)}]{misra_zenos_1977}%
  \BibitemOpen
  \bibfield  {author} {\bibinfo {author} {\bibfnamefont {B.}~\bibnamefont
  {Misra}}\ and\ \bibinfo {author} {\bibfnamefont {E.~C.~G.}\ \bibnamefont
  {Sudarshan}},\ }\href {\doibase 10.1063/1.523304} {\bibfield  {journal}
  {\bibinfo  {journal} {J. Math. Phys.}\ }\textbf {\bibinfo {volume} {18}},\
  \bibinfo {pages} {756} (\bibinfo {year} {1977})}\BibitemShut {NoStop}%
\bibitem [{\citenamefont {Itano}\ \emph {et~al.}(1990)\citenamefont {Itano},
  \citenamefont {Heinzen}, \citenamefont {Bollinger},\ and\ \citenamefont
  {Wineland}}]{itano_quantum_1990}%
  \BibitemOpen
  \bibfield  {author} {\bibinfo {author} {\bibfnamefont {W.~M.}\ \bibnamefont
  {Itano}}, \bibinfo {author} {\bibfnamefont {D.~J.}\ \bibnamefont {Heinzen}},
  \bibinfo {author} {\bibfnamefont {J.~J.}\ \bibnamefont {Bollinger}}, \ and\
  \bibinfo {author} {\bibfnamefont {D.~J.}\ \bibnamefont {Wineland}},\ }\href
  {\doibase 10.1103/PhysRevA.41.2295} {\bibfield  {journal} {\bibinfo
  {journal} {Phys. Rev. A}\ }\textbf {\bibinfo {volume} {41}},\ \bibinfo
  {pages} {2295} (\bibinfo {year} {1990})}\BibitemShut {NoStop}%
\bibitem [{\citenamefont {Wolters}\ \emph {et~al.}(2013)\citenamefont
  {Wolters}, \citenamefont {Strau{\ss}}, \citenamefont {Schoenfeld},\ and\
  \citenamefont {Benson}}]{wolters_quantum_2013}%
  \BibitemOpen
  \bibfield  {author} {\bibinfo {author} {\bibfnamefont {J.}~\bibnamefont
  {Wolters}}, \bibinfo {author} {\bibfnamefont {M.}~\bibnamefont {Strau{\ss}}},
  \bibinfo {author} {\bibfnamefont {R.~S.}\ \bibnamefont {Schoenfeld}}, \ and\
  \bibinfo {author} {\bibfnamefont {O.}~\bibnamefont {Benson}},\ }\href
  {\doibase 10.1103/PhysRevA.88.020101} {\bibfield  {journal} {\bibinfo
  {journal} {Phys. Rev. A}\ }\textbf {\bibinfo {volume} {88}},\ \bibinfo
  {pages} {020101} (\bibinfo {year} {2013})}\BibitemShut {NoStop}%
\bibitem [{\citenamefont {Kakuyanagi}\ \emph {et~al.}(2015)\citenamefont
  {Kakuyanagi}, \citenamefont {Baba}, \citenamefont {Matsuzaki}, \citenamefont
  {Nakano}, \citenamefont {Saito},\ and\ \citenamefont
  {Semba}}]{kakuyanagi_observation_2015}%
  \BibitemOpen
  \bibfield  {author} {\bibinfo {author} {\bibfnamefont {K.}~\bibnamefont
  {Kakuyanagi}}, \bibinfo {author} {\bibfnamefont {T.}~\bibnamefont {Baba}},
  \bibinfo {author} {\bibfnamefont {Y.}~\bibnamefont {Matsuzaki}}, \bibinfo
  {author} {\bibfnamefont {H.}~\bibnamefont {Nakano}}, \bibinfo {author}
  {\bibfnamefont {S.}~\bibnamefont {Saito}}, \ and\ \bibinfo {author}
  {\bibfnamefont {K.}~\bibnamefont {Semba}},\ }\href {\doibase
  10.1088/1367-2630/17/6/063035} {\bibfield  {journal} {\bibinfo  {journal}
  {New J. Phys.}\ }\textbf {\bibinfo {volume} {17}},\ \bibinfo {pages} {063035}
  (\bibinfo {year} {2015})}\BibitemShut {NoStop}%
\bibitem [{\citenamefont {Peise}\ \emph {et~al.}(2015)\citenamefont {Peise},
  \citenamefont {L{\"u}cke}, \citenamefont {Pezz{\'e}}, \citenamefont
  {Deuretzbacher}, \citenamefont {Ertmer}, \citenamefont {Arlt}, \citenamefont
  {Smerzi}, \citenamefont {Santos},\ and\ \citenamefont
  {Klempt}}]{peise_interactionfree_2015}%
  \BibitemOpen
  \bibfield  {author} {\bibinfo {author} {\bibfnamefont {J.}~\bibnamefont
  {Peise}}, \bibinfo {author} {\bibfnamefont {B.}~\bibnamefont {L{\"u}cke}},
  \bibinfo {author} {\bibfnamefont {L.}~\bibnamefont {Pezz{\'e}}}, \bibinfo
  {author} {\bibfnamefont {F.}~\bibnamefont {Deuretzbacher}}, \bibinfo {author}
  {\bibfnamefont {W.}~\bibnamefont {Ertmer}}, \bibinfo {author} {\bibfnamefont
  {J.}~\bibnamefont {Arlt}}, \bibinfo {author} {\bibfnamefont {A.}~\bibnamefont
  {Smerzi}}, \bibinfo {author} {\bibfnamefont {L.}~\bibnamefont {Santos}}, \
  and\ \bibinfo {author} {\bibfnamefont {C.}~\bibnamefont {Klempt}},\ }\href
  {\doibase 10.1038/ncomms7811} {\bibfield  {journal} {\bibinfo  {journal} {Nat
  Commun}\ }\textbf {\bibinfo {volume} {6}},\ \bibinfo {pages} {6811} (\bibinfo
  {year} {2015})}\BibitemShut {NoStop}%
\bibitem [{\citenamefont {Slichter}\ \emph {et~al.}(2015)\citenamefont
  {Slichter}, \citenamefont {M{\"u}ller}, \citenamefont {Vijay}, \citenamefont
  {Weber}, \citenamefont {Blais},\ and\ \citenamefont
  {Siddiqi}}]{slichter_quantum_2015}%
  \BibitemOpen
  \bibfield  {author} {\bibinfo {author} {\bibfnamefont {D.~H.}\ \bibnamefont
  {Slichter}}, \bibinfo {author} {\bibfnamefont {C.}~\bibnamefont
  {M{\"u}ller}}, \bibinfo {author} {\bibfnamefont {R.}~\bibnamefont {Vijay}},
  \bibinfo {author} {\bibfnamefont {S.~J.}\ \bibnamefont {Weber}}, \bibinfo
  {author} {\bibfnamefont {A.}~\bibnamefont {Blais}}, \ and\ \bibinfo {author}
  {\bibfnamefont {I.}~\bibnamefont {Siddiqi}},\ }\href
  {http://arxiv.org/abs/1512.04006} {\bibfield  {journal} {\bibinfo  {journal}
  {arXiv:1512.04006}\ } (\bibinfo {year} {2015})}\BibitemShut {NoStop}%
\bibitem [{\citenamefont {Patil}\ \emph {et~al.}(2015)\citenamefont {Patil},
  \citenamefont {Chakram},\ and\ \citenamefont
  {Vengalattore}}]{patil_measurementinduced_2015}%
  \BibitemOpen
  \bibfield  {author} {\bibinfo {author} {\bibfnamefont {Y.}~\bibnamefont
  {Patil}}, \bibinfo {author} {\bibfnamefont {S.}~\bibnamefont {Chakram}}, \
  and\ \bibinfo {author} {\bibfnamefont {M.}~\bibnamefont {Vengalattore}},\
  }\href {\doibase 10.1103/PhysRevLett.115.140402} {\bibfield  {journal}
  {\bibinfo  {journal} {Phys. Rev. Lett.}\ }\textbf {\bibinfo {volume} {115}},\
  \bibinfo {pages} {140402} (\bibinfo {year} {2015})}\BibitemShut {NoStop}%
\bibitem [{\citenamefont {Wilkinson}\ \emph {et~al.}(1997)\citenamefont
  {Wilkinson}, \citenamefont {Bharucha}, \citenamefont {Fischer}, \citenamefont
  {Madison}, \citenamefont {Morrow}, \citenamefont {Niu}, \citenamefont
  {Sundaram},\ and\ \citenamefont {Raizen}}]{wilkinson_experimental_1997}%
  \BibitemOpen
  \bibfield  {author} {\bibinfo {author} {\bibfnamefont {S.~R.}\ \bibnamefont
  {Wilkinson}}, \bibinfo {author} {\bibfnamefont {C.~F.}\ \bibnamefont
  {Bharucha}}, \bibinfo {author} {\bibfnamefont {M.~C.}\ \bibnamefont
  {Fischer}}, \bibinfo {author} {\bibfnamefont {K.~W.}\ \bibnamefont
  {Madison}}, \bibinfo {author} {\bibfnamefont {P.~R.}\ \bibnamefont {Morrow}},
  \bibinfo {author} {\bibfnamefont {Q.}~\bibnamefont {Niu}}, \bibinfo {author}
  {\bibfnamefont {B.}~\bibnamefont {Sundaram}}, \ and\ \bibinfo {author}
  {\bibfnamefont {M.~G.}\ \bibnamefont {Raizen}},\ }\href {\doibase
  10.1038/42418} {\bibfield  {journal} {\bibinfo  {journal} {Nature}\ }\textbf
  {\bibinfo {volume} {387}},\ \bibinfo {pages} {575} (\bibinfo {year}
  {1997})}\BibitemShut {NoStop}%
\bibitem [{\citenamefont {Fischer}\ \emph {et~al.}(2001)\citenamefont
  {Fischer}, \citenamefont {Guti{\'e}rrez-Medina},\ and\ \citenamefont
  {Raizen}}]{fischer_observation_2001}%
  \BibitemOpen
  \bibfield  {author} {\bibinfo {author} {\bibfnamefont {M.~C.}\ \bibnamefont
  {Fischer}}, \bibinfo {author} {\bibfnamefont {B.}~\bibnamefont
  {Guti{\'e}rrez-Medina}}, \ and\ \bibinfo {author} {\bibfnamefont {M.~G.}\
  \bibnamefont {Raizen}},\ }\href {\doibase 10.1103/PhysRevLett.87.040402}
  {\bibfield  {journal} {\bibinfo  {journal} {Phys. Rev. Lett.}\ }\textbf
  {\bibinfo {volume} {87}},\ \bibinfo {pages} {040402} (\bibinfo {year}
  {2001})}\BibitemShut {NoStop}%
\bibitem [{\citenamefont {Facchi}\ and\ \citenamefont
  {Pascazio}(2002)}]{facchi_quantum_2002}%
  \BibitemOpen
  \bibfield  {author} {\bibinfo {author} {\bibfnamefont {P.}~\bibnamefont
  {Facchi}}\ and\ \bibinfo {author} {\bibfnamefont {S.}~\bibnamefont
  {Pascazio}},\ }\href {\doibase 10.1103/PhysRevLett.89.080401} {\bibfield
  {journal} {\bibinfo  {journal} {Phys. Rev. Lett.}\ }\textbf {\bibinfo
  {volume} {89}},\ \bibinfo {pages} {080401} (\bibinfo {year}
  {2002})}\BibitemShut {NoStop}%
\bibitem [{\citenamefont {Facchi}\ \emph {et~al.}(2004)\citenamefont {Facchi},
  \citenamefont {Lidar},\ and\ \citenamefont
  {Pascazio}}]{facchi_unification_2004}%
  \BibitemOpen
  \bibfield  {author} {\bibinfo {author} {\bibfnamefont {P.}~\bibnamefont
  {Facchi}}, \bibinfo {author} {\bibfnamefont {D.~A.}\ \bibnamefont {Lidar}}, \
  and\ \bibinfo {author} {\bibfnamefont {S.}~\bibnamefont {Pascazio}},\ }\href
  {\doibase 10.1103/PhysRevA.69.032314} {\bibfield  {journal} {\bibinfo
  {journal} {Phys. Rev. A}\ }\textbf {\bibinfo {volume} {69}},\ \bibinfo
  {pages} {032314} (\bibinfo {year} {2004})}\BibitemShut {NoStop}%
\bibitem [{\citenamefont {Viola}\ and\ \citenamefont
  {Lloyd}(1998)}]{viola_dynamical_1998}%
  \BibitemOpen
  \bibfield  {author} {\bibinfo {author} {\bibfnamefont {L.}~\bibnamefont
  {Viola}}\ and\ \bibinfo {author} {\bibfnamefont {S.}~\bibnamefont {Lloyd}},\
  }\href {\doibase 10.1103/PhysRevA.58.2733} {\bibfield  {journal} {\bibinfo
  {journal} {Phys. Rev. A}\ }\textbf {\bibinfo {volume} {58}},\ \bibinfo
  {pages} {2733} (\bibinfo {year} {1998})}\BibitemShut {NoStop}%
\bibitem [{\citenamefont {Dhar}\ \emph {et~al.}(2006)\citenamefont {Dhar},
  \citenamefont {Grover},\ and\ \citenamefont {Roy}}]{dhar_preserving_2006}%
  \BibitemOpen
  \bibfield  {author} {\bibinfo {author} {\bibfnamefont {D.}~\bibnamefont
  {Dhar}}, \bibinfo {author} {\bibfnamefont {L.~K.}\ \bibnamefont {Grover}}, \
  and\ \bibinfo {author} {\bibfnamefont {S.~M.}\ \bibnamefont {Roy}},\ }\href
  {\doibase 10.1103/PhysRevLett.96.100405} {\bibfield  {journal} {\bibinfo
  {journal} {Phys. Rev. Lett.}\ }\textbf {\bibinfo {volume} {96}},\ \bibinfo
  {pages} {100405} (\bibinfo {year} {2006})}\BibitemShut {NoStop}%
\bibitem [{\citenamefont {Zheng}\ \emph {et~al.}(2013)\citenamefont {Zheng},
  \citenamefont {Xu}, \citenamefont {Peng}, \citenamefont {Zhou}, \citenamefont
  {Du},\ and\ \citenamefont {Sun}}]{zheng_experimental_2013}%
  \BibitemOpen
  \bibfield  {author} {\bibinfo {author} {\bibfnamefont {W.}~\bibnamefont
  {Zheng}}, \bibinfo {author} {\bibfnamefont {D.~Z.}\ \bibnamefont {Xu}},
  \bibinfo {author} {\bibfnamefont {X.}~\bibnamefont {Peng}}, \bibinfo {author}
  {\bibfnamefont {X.}~\bibnamefont {Zhou}}, \bibinfo {author} {\bibfnamefont
  {J.}~\bibnamefont {Du}}, \ and\ \bibinfo {author} {\bibfnamefont {C.~P.}\
  \bibnamefont {Sun}},\ }\href {\doibase 10.1103/PhysRevA.87.032112} {\bibfield
   {journal} {\bibinfo  {journal} {Phys. Rev. A}\ }\textbf {\bibinfo {volume}
  {87}},\ \bibinfo {pages} {032112} (\bibinfo {year} {2013})}\BibitemShut
  {NoStop}%
\bibitem [{\citenamefont {Singh}\ \emph {et~al.}(2014)\citenamefont {Singh},
  \citenamefont {{Arvind}},\ and\ \citenamefont
  {Dorai}}]{singh_experimental_2014}%
  \BibitemOpen
  \bibfield  {author} {\bibinfo {author} {\bibfnamefont {H.}~\bibnamefont
  {Singh}}, \bibinfo {author} {\bibnamefont {{Arvind}}}, \ and\ \bibinfo
  {author} {\bibfnamefont {K.}~\bibnamefont {Dorai}},\ }\href {\doibase
  10.1103/PhysRevA.90.052329} {\bibfield  {journal} {\bibinfo  {journal} {Phys.
  Rev. A}\ }\textbf {\bibinfo {volume} {90}},\ \bibinfo {pages} {052329}
  (\bibinfo {year} {2014})}\BibitemShut {NoStop}%
\bibitem [{\citenamefont {Zhong}\ \emph {et~al.}(2015)\citenamefont {Zhong},
  \citenamefont {Hedges}, \citenamefont {Ahlefeldt}, \citenamefont
  {Bartholomew}, \citenamefont {Beavan}, \citenamefont {Wittig}, \citenamefont
  {Longdell},\ and\ \citenamefont {Sellars}}]{zhong_optically_2015}%
  \BibitemOpen
  \bibfield  {author} {\bibinfo {author} {\bibfnamefont {M.}~\bibnamefont
  {Zhong}}, \bibinfo {author} {\bibfnamefont {M.~P.}\ \bibnamefont {Hedges}},
  \bibinfo {author} {\bibfnamefont {R.~L.}\ \bibnamefont {Ahlefeldt}}, \bibinfo
  {author} {\bibfnamefont {J.~G.}\ \bibnamefont {Bartholomew}}, \bibinfo
  {author} {\bibfnamefont {S.~E.}\ \bibnamefont {Beavan}}, \bibinfo {author}
  {\bibfnamefont {S.~M.}\ \bibnamefont {Wittig}}, \bibinfo {author}
  {\bibfnamefont {J.~J.}\ \bibnamefont {Longdell}}, \ and\ \bibinfo {author}
  {\bibfnamefont {M.~J.}\ \bibnamefont {Sellars}},\ }\href {\doibase
  10.1038/nature14025} {\bibfield  {journal} {\bibinfo  {journal} {Nature}\
  }\textbf {\bibinfo {volume} {517}},\ \bibinfo {pages} {177} (\bibinfo {year}
  {2015})}\BibitemShut {NoStop}%
\bibitem [{\citenamefont {Maurer}\ \emph {et~al.}(2012)\citenamefont {Maurer},
  \citenamefont {Kucsko}, \citenamefont {Latta}, \citenamefont {Jiang},
  \citenamefont {Yao}, \citenamefont {Bennett}, \citenamefont {Pastawski},
  \citenamefont {Hunger}, \citenamefont {Chisholm}, \citenamefont {Markham},
  \citenamefont {Twitchen}, \citenamefont {Cirac},\ and\ \citenamefont
  {Lukin}}]{maurer_roomtemperature_2012}%
  \BibitemOpen
  \bibfield  {author} {\bibinfo {author} {\bibfnamefont {P.~C.}\ \bibnamefont
  {Maurer}}, \bibinfo {author} {\bibfnamefont {G.}~\bibnamefont {Kucsko}},
  \bibinfo {author} {\bibfnamefont {C.}~\bibnamefont {Latta}}, \bibinfo
  {author} {\bibfnamefont {L.}~\bibnamefont {Jiang}}, \bibinfo {author}
  {\bibfnamefont {N.~Y.}\ \bibnamefont {Yao}}, \bibinfo {author} {\bibfnamefont
  {S.~D.}\ \bibnamefont {Bennett}}, \bibinfo {author} {\bibfnamefont
  {F.}~\bibnamefont {Pastawski}}, \bibinfo {author} {\bibfnamefont
  {D.}~\bibnamefont {Hunger}}, \bibinfo {author} {\bibfnamefont
  {N.}~\bibnamefont {Chisholm}}, \bibinfo {author} {\bibfnamefont
  {M.}~\bibnamefont {Markham}}, \bibinfo {author} {\bibfnamefont {D.~J.}\
  \bibnamefont {Twitchen}}, \bibinfo {author} {\bibfnamefont {J.~I.}\
  \bibnamefont {Cirac}}, \ and\ \bibinfo {author} {\bibfnamefont {M.~D.}\
  \bibnamefont {Lukin}},\ }\href {\doibase 10.1126/science.1220513} {\bibfield
  {journal} {\bibinfo  {journal} {Science}\ }\textbf {\bibinfo {volume}
  {336}},\ \bibinfo {pages} {1283} (\bibinfo {year} {2012})}\BibitemShut
  {NoStop}%
\bibitem [{\citenamefont {{de Lange}}\ \emph {et~al.}(2010)\citenamefont {{de
  Lange}}, \citenamefont {Wang}, \citenamefont {Rist{\`e}}, \citenamefont
  {Dobrovitski},\ and\ \citenamefont {Hanson}}]{lange_universal_2010}%
  \BibitemOpen
  \bibfield  {author} {\bibinfo {author} {\bibfnamefont {G.}~\bibnamefont {{de
  Lange}}}, \bibinfo {author} {\bibfnamefont {Z.~H.}\ \bibnamefont {Wang}},
  \bibinfo {author} {\bibfnamefont {D.}~\bibnamefont {Rist{\`e}}}, \bibinfo
  {author} {\bibfnamefont {V.~V.}\ \bibnamefont {Dobrovitski}}, \ and\ \bibinfo
  {author} {\bibfnamefont {R.}~\bibnamefont {Hanson}},\ }\href {\doibase
  10.1126/science.1192739} {\bibfield  {journal} {\bibinfo  {journal}
  {Science}\ }\textbf {\bibinfo {volume} {330}},\ \bibinfo {pages} {60}
  (\bibinfo {year} {2010})}\BibitemShut {NoStop}%
\bibitem [{\citenamefont {Leghtas}\ \emph {et~al.}(2015)\citenamefont
  {Leghtas}, \citenamefont {Touzard}, \citenamefont {Pop}, \citenamefont {Kou},
  \citenamefont {Vlastakis}, \citenamefont {Petrenko}, \citenamefont {Sliwa},
  \citenamefont {Narla}, \citenamefont {Shankar}, \citenamefont {Hatridge},
  \citenamefont {Reagor}, \citenamefont {Frunzio}, \citenamefont {Schoelkopf},
  \citenamefont {Mirrahimi},\ and\ \citenamefont
  {Devoret}}]{leghtas_confining_2015}%
  \BibitemOpen
  \bibfield  {author} {\bibinfo {author} {\bibfnamefont {Z.}~\bibnamefont
  {Leghtas}}, \bibinfo {author} {\bibfnamefont {S.}~\bibnamefont {Touzard}},
  \bibinfo {author} {\bibfnamefont {I.~M.}\ \bibnamefont {Pop}}, \bibinfo
  {author} {\bibfnamefont {A.}~\bibnamefont {Kou}}, \bibinfo {author}
  {\bibfnamefont {B.}~\bibnamefont {Vlastakis}}, \bibinfo {author}
  {\bibfnamefont {A.}~\bibnamefont {Petrenko}}, \bibinfo {author}
  {\bibfnamefont {K.~M.}\ \bibnamefont {Sliwa}}, \bibinfo {author}
  {\bibfnamefont {A.}~\bibnamefont {Narla}}, \bibinfo {author} {\bibfnamefont
  {S.}~\bibnamefont {Shankar}}, \bibinfo {author} {\bibfnamefont {M.~J.}\
  \bibnamefont {Hatridge}}, \bibinfo {author} {\bibfnamefont {M.}~\bibnamefont
  {Reagor}}, \bibinfo {author} {\bibfnamefont {L.}~\bibnamefont {Frunzio}},
  \bibinfo {author} {\bibfnamefont {R.~J.}\ \bibnamefont {Schoelkopf}},
  \bibinfo {author} {\bibfnamefont {M.}~\bibnamefont {Mirrahimi}}, \ and\
  \bibinfo {author} {\bibfnamefont {M.~H.}\ \bibnamefont {Devoret}},\ }\href
  {\doibase 10.1126/science.aaa2085} {\bibfield  {journal} {\bibinfo  {journal}
  {Science}\ }\textbf {\bibinfo {volume} {347}},\ \bibinfo {pages} {853}
  (\bibinfo {year} {2015})}\BibitemShut {NoStop}%
\bibitem [{\citenamefont {Bretheau}\ \emph {et~al.}(2015)\citenamefont
  {Bretheau}, \citenamefont {Campagne-Ibarcq}, \citenamefont {Flurin},
  \citenamefont {Mallet},\ and\ \citenamefont {Huard}}]{bretheau_quantum_2015}%
  \BibitemOpen
  \bibfield  {author} {\bibinfo {author} {\bibfnamefont {L.}~\bibnamefont
  {Bretheau}}, \bibinfo {author} {\bibfnamefont {P.}~\bibnamefont
  {Campagne-Ibarcq}}, \bibinfo {author} {\bibfnamefont {E.}~\bibnamefont
  {Flurin}}, \bibinfo {author} {\bibfnamefont {F.}~\bibnamefont {Mallet}}, \
  and\ \bibinfo {author} {\bibfnamefont {B.}~\bibnamefont {Huard}},\ }\href
  {\doibase 10.1126/science.1259345} {\bibfield  {journal} {\bibinfo  {journal}
  {Science}\ }\textbf {\bibinfo {volume} {348}},\ \bibinfo {pages} {776}
  (\bibinfo {year} {2015})}\BibitemShut {NoStop}%
\bibitem [{\citenamefont {Signoles}\ \emph {et~al.}(2014)\citenamefont
  {Signoles}, \citenamefont {Facon}, \citenamefont {Grosso}, \citenamefont
  {Dotsenko}, \citenamefont {Haroche}, \citenamefont {Raimond}, \citenamefont
  {Brune},\ and\ \citenamefont {Gleyzes}}]{signoles_confined_2014}%
  \BibitemOpen
  \bibfield  {author} {\bibinfo {author} {\bibfnamefont {A.}~\bibnamefont
  {Signoles}}, \bibinfo {author} {\bibfnamefont {A.}~\bibnamefont {Facon}},
  \bibinfo {author} {\bibfnamefont {D.}~\bibnamefont {Grosso}}, \bibinfo
  {author} {\bibfnamefont {I.}~\bibnamefont {Dotsenko}}, \bibinfo {author}
  {\bibfnamefont {S.}~\bibnamefont {Haroche}}, \bibinfo {author} {\bibfnamefont
  {J.-M.}\ \bibnamefont {Raimond}}, \bibinfo {author} {\bibfnamefont
  {M.}~\bibnamefont {Brune}}, \ and\ \bibinfo {author} {\bibfnamefont
  {S.}~\bibnamefont {Gleyzes}},\ }\href {\doibase 10.1038/nphys3076} {\bibfield
   {journal} {\bibinfo  {journal} {Nat Phys}\ }\textbf {\bibinfo {volume}
  {10}},\ \bibinfo {pages} {715} (\bibinfo {year} {2014})}\BibitemShut
  {NoStop}%
\bibitem [{\citenamefont {Barontini}\ \emph {et~al.}(2015)\citenamefont
  {Barontini}, \citenamefont {Hohmann}, \citenamefont {Haas}, \citenamefont
  {Est{\`e}ve},\ and\ \citenamefont {Reichel}}]{barontini_deterministic_2015}%
  \BibitemOpen
  \bibfield  {author} {\bibinfo {author} {\bibfnamefont {G.}~\bibnamefont
  {Barontini}}, \bibinfo {author} {\bibfnamefont {L.}~\bibnamefont {Hohmann}},
  \bibinfo {author} {\bibfnamefont {F.}~\bibnamefont {Haas}}, \bibinfo {author}
  {\bibfnamefont {J.}~\bibnamefont {Est{\`e}ve}}, \ and\ \bibinfo {author}
  {\bibfnamefont {J.}~\bibnamefont {Reichel}},\ }\href {\doibase
  10.1126/science.aaa0754} {\bibfield  {journal} {\bibinfo  {journal}
  {Science}\ }\textbf {\bibinfo {volume} {349}},\ \bibinfo {pages} {1317}
  (\bibinfo {year} {2015})}\BibitemShut {NoStop}%
\bibitem [{\citenamefont {Sch{\"a}fer}\ \emph {et~al.}(2014)\citenamefont
  {Sch{\"a}fer}, \citenamefont {Herrera}, \citenamefont {Cherukattil},
  \citenamefont {Lovecchio}, \citenamefont {Cataliotti}, \citenamefont
  {Caruso},\ and\ \citenamefont {Smerzi}}]{schafer_experimental_2014}%
  \BibitemOpen
  \bibfield  {author} {\bibinfo {author} {\bibfnamefont {F.}~\bibnamefont
  {Sch{\"a}fer}}, \bibinfo {author} {\bibfnamefont {I.}~\bibnamefont
  {Herrera}}, \bibinfo {author} {\bibfnamefont {S.}~\bibnamefont
  {Cherukattil}}, \bibinfo {author} {\bibfnamefont {C.}~\bibnamefont
  {Lovecchio}}, \bibinfo {author} {\bibfnamefont {F.~S.}\ \bibnamefont
  {Cataliotti}}, \bibinfo {author} {\bibfnamefont {F.}~\bibnamefont {Caruso}},
  \ and\ \bibinfo {author} {\bibfnamefont {A.}~\bibnamefont {Smerzi}},\ }\href
  {\doibase 10.1038/ncomms4194} {\bibfield  {journal} {\bibinfo  {journal} {Nat
  Commun}\ }\textbf {\bibinfo {volume} {5}} (\bibinfo {year} {2014}),\
  10.1038/ncomms4194}\BibitemShut {NoStop}%
\bibitem [{\citenamefont {Cramer}\ \emph {et~al.}(2015)\citenamefont {Cramer},
  \citenamefont {Kalb}, \citenamefont {Rol}, \citenamefont {Hensen},
  \citenamefont {Blok}, \citenamefont {Markham}, \citenamefont {Twitchen},
  \citenamefont {Hanson},\ and\ \citenamefont
  {Taminiau}}]{cramer_repeated_2015}%
  \BibitemOpen
  \bibfield  {author} {\bibinfo {author} {\bibfnamefont {J.}~\bibnamefont
  {Cramer}}, \bibinfo {author} {\bibfnamefont {N.}~\bibnamefont {Kalb}},
  \bibinfo {author} {\bibfnamefont {M.~A.}\ \bibnamefont {Rol}}, \bibinfo
  {author} {\bibfnamefont {B.}~\bibnamefont {Hensen}}, \bibinfo {author}
  {\bibfnamefont {M.~S.}\ \bibnamefont {Blok}}, \bibinfo {author}
  {\bibfnamefont {M.}~\bibnamefont {Markham}}, \bibinfo {author} {\bibfnamefont
  {D.~J.}\ \bibnamefont {Twitchen}}, \bibinfo {author} {\bibfnamefont
  {R.}~\bibnamefont {Hanson}}, \ and\ \bibinfo {author} {\bibfnamefont {T.~H.}\
  \bibnamefont {Taminiau}},\ }\href {http://arxiv.org/abs/1508.01388}
  {\bibfield  {journal} {\bibinfo  {journal} {arXiv:1508.01388}\ } (\bibinfo
  {year} {2015})}\BibitemShut {NoStop}%
\bibitem [{\citenamefont {Bloembergen}(1949)}]{bloembergen_interaction_1949}%
  \BibitemOpen
  \bibfield  {author} {\bibinfo {author} {\bibfnamefont {N.}~\bibnamefont
  {Bloembergen}},\ }\href {\doibase 10.1016/0031-8914(49)90114-7} {\bibfield
  {journal} {\bibinfo  {journal} {Physica}\ }\textbf {\bibinfo {volume} {15}},\
  \bibinfo {pages} {386} (\bibinfo {year} {1949})}\BibitemShut {NoStop}%
\bibitem [{\citenamefont {Robledo}\ \emph {et~al.}(2011)\citenamefont
  {Robledo}, \citenamefont {Childress}, \citenamefont {Bernien}, \citenamefont
  {Hensen}, \citenamefont {Alkemade},\ and\ \citenamefont
  {Hanson}}]{robledo_highfidelity_2011}%
  \BibitemOpen
  \bibfield  {author} {\bibinfo {author} {\bibfnamefont {L.}~\bibnamefont
  {Robledo}}, \bibinfo {author} {\bibfnamefont {L.}~\bibnamefont {Childress}},
  \bibinfo {author} {\bibfnamefont {H.}~\bibnamefont {Bernien}}, \bibinfo
  {author} {\bibfnamefont {B.}~\bibnamefont {Hensen}}, \bibinfo {author}
  {\bibfnamefont {P.~F.~A.}\ \bibnamefont {Alkemade}}, \ and\ \bibinfo {author}
  {\bibfnamefont {R.}~\bibnamefont {Hanson}},\ }\href {\doibase
  10.1038/nature10401} {\bibfield  {journal} {\bibinfo  {journal} {Nature}\
  }\textbf {\bibinfo {volume} {477}},\ \bibinfo {pages} {574} (\bibinfo {year}
  {2011})}\BibitemShut {NoStop}%
\bibitem [{\citenamefont {Taminiau}\ \emph {et~al.}(2014)\citenamefont
  {Taminiau}, \citenamefont {Cramer}, \citenamefont {{van der Sar}},
  \citenamefont {Dobrovitski},\ and\ \citenamefont
  {Hanson}}]{taminiau_universal_2014}%
  \BibitemOpen
  \bibfield  {author} {\bibinfo {author} {\bibfnamefont {T.~H.}\ \bibnamefont
  {Taminiau}}, \bibinfo {author} {\bibfnamefont {J.}~\bibnamefont {Cramer}},
  \bibinfo {author} {\bibfnamefont {T.}~\bibnamefont {{van der Sar}}}, \bibinfo
  {author} {\bibfnamefont {V.~V.}\ \bibnamefont {Dobrovitski}}, \ and\ \bibinfo
  {author} {\bibfnamefont {R.}~\bibnamefont {Hanson}},\ }\href {\doibase
  10.1038/nnano.2014.2} {\bibfield  {journal} {\bibinfo  {journal} {Nat Nano}\
  }\textbf {\bibinfo {volume} {9}},\ \bibinfo {pages} {171} (\bibinfo {year}
  {2014})}\BibitemShut {NoStop}%
\bibitem [{\citenamefont {Waldherr}\ \emph {et~al.}(2014)\citenamefont
  {Waldherr}, \citenamefont {Wang}, \citenamefont {Zaiser}, \citenamefont
  {Jamali}, \citenamefont {Schulte-Herbr{\"u}ggen}, \citenamefont {Abe},
  \citenamefont {Ohshima}, \citenamefont {Isoya}, \citenamefont {Du},
  \citenamefont {Neumann},\ and\ \citenamefont
  {Wrachtrup}}]{waldherr_quantum_2014}%
  \BibitemOpen
  \bibfield  {author} {\bibinfo {author} {\bibfnamefont {G.}~\bibnamefont
  {Waldherr}}, \bibinfo {author} {\bibfnamefont {Y.}~\bibnamefont {Wang}},
  \bibinfo {author} {\bibfnamefont {S.}~\bibnamefont {Zaiser}}, \bibinfo
  {author} {\bibfnamefont {M.}~\bibnamefont {Jamali}}, \bibinfo {author}
  {\bibfnamefont {T.}~\bibnamefont {Schulte-Herbr{\"u}ggen}}, \bibinfo {author}
  {\bibfnamefont {H.}~\bibnamefont {Abe}}, \bibinfo {author} {\bibfnamefont
  {T.}~\bibnamefont {Ohshima}}, \bibinfo {author} {\bibfnamefont
  {J.}~\bibnamefont {Isoya}}, \bibinfo {author} {\bibfnamefont {J.~F.}\
  \bibnamefont {Du}}, \bibinfo {author} {\bibfnamefont {P.}~\bibnamefont
  {Neumann}}, \ and\ \bibinfo {author} {\bibfnamefont {J.}~\bibnamefont
  {Wrachtrup}},\ }\href {\doibase 10.1038/nature12919} {\bibfield  {journal}
  {\bibinfo  {journal} {Nature}\ }\textbf {\bibinfo {volume} {506}},\ \bibinfo
  {pages} {204} (\bibinfo {year} {2014})}\BibitemShut {NoStop}%
\bibitem [{\citenamefont {Dr{\'e}au}\ \emph {et~al.}(2013)\citenamefont
  {Dr{\'e}au}, \citenamefont {Spinicelli}, \citenamefont {Maze}, \citenamefont
  {Roch},\ and\ \citenamefont {Jacques}}]{dreau_single-shot_2013}%
  \BibitemOpen
  \bibfield  {author} {\bibinfo {author} {\bibfnamefont {A.}~\bibnamefont
  {Dr{\'e}au}}, \bibinfo {author} {\bibfnamefont {P.}~\bibnamefont
  {Spinicelli}}, \bibinfo {author} {\bibfnamefont {J.~R.}\ \bibnamefont
  {Maze}}, \bibinfo {author} {\bibfnamefont {J.-F.}\ \bibnamefont {Roch}}, \
  and\ \bibinfo {author} {\bibfnamefont {V.}~\bibnamefont {Jacques}},\ }\href
  {\doibase 10.1103/PhysRevLett.110.060502} {\bibfield  {journal} {\bibinfo
  {journal} {Phys. Rev. Lett.}\ }\textbf {\bibinfo {volume} {110}},\ \bibinfo
  {pages} {060502} (\bibinfo {year} {2013})}\BibitemShut {NoStop}%
\bibitem [{\citenamefont {Jiang}\ \emph {et~al.}(2009)\citenamefont {Jiang},
  \citenamefont {Hodges}, \citenamefont {Maze}, \citenamefont {Maurer},
  \citenamefont {Taylor}, \citenamefont {Cory}, \citenamefont {Hemmer},
  \citenamefont {Walsworth}, \citenamefont {Yacoby}, \citenamefont {Zibrov},\
  and\ \citenamefont {Lukin}}]{jiang_repetitive_2009}%
  \BibitemOpen
  \bibfield  {author} {\bibinfo {author} {\bibfnamefont {L.}~\bibnamefont
  {Jiang}}, \bibinfo {author} {\bibfnamefont {J.~S.}\ \bibnamefont {Hodges}},
  \bibinfo {author} {\bibfnamefont {J.~R.}\ \bibnamefont {Maze}}, \bibinfo
  {author} {\bibfnamefont {P.}~\bibnamefont {Maurer}}, \bibinfo {author}
  {\bibfnamefont {J.~M.}\ \bibnamefont {Taylor}}, \bibinfo {author}
  {\bibfnamefont {D.~G.}\ \bibnamefont {Cory}}, \bibinfo {author}
  {\bibfnamefont {P.~R.}\ \bibnamefont {Hemmer}}, \bibinfo {author}
  {\bibfnamefont {R.~L.}\ \bibnamefont {Walsworth}}, \bibinfo {author}
  {\bibfnamefont {A.}~\bibnamefont {Yacoby}}, \bibinfo {author} {\bibfnamefont
  {A.~S.}\ \bibnamefont {Zibrov}}, \ and\ \bibinfo {author} {\bibfnamefont
  {M.~D.}\ \bibnamefont {Lukin}},\ }\href {\doibase 10.1126/science.1176496}
  {\bibfield  {journal} {\bibinfo  {journal} {Science}\ }\textbf {\bibinfo
  {volume} {326}},\ \bibinfo {pages} {267} (\bibinfo {year}
  {2009})}\BibitemShut {NoStop}%
\bibitem [{\citenamefont {Neumann}\ \emph {et~al.}(2010)\citenamefont
  {Neumann}, \citenamefont {Beck}, \citenamefont {Steiner}, \citenamefont
  {Rempp}, \citenamefont {Fedder}, \citenamefont {Hemmer}, \citenamefont
  {Wrachtrup},\ and\ \citenamefont {Jelezko}}]{neumann_single-shot_2010}%
  \BibitemOpen
  \bibfield  {author} {\bibinfo {author} {\bibfnamefont {P.}~\bibnamefont
  {Neumann}}, \bibinfo {author} {\bibfnamefont {J.}~\bibnamefont {Beck}},
  \bibinfo {author} {\bibfnamefont {M.}~\bibnamefont {Steiner}}, \bibinfo
  {author} {\bibfnamefont {F.}~\bibnamefont {Rempp}}, \bibinfo {author}
  {\bibfnamefont {H.}~\bibnamefont {Fedder}}, \bibinfo {author} {\bibfnamefont
  {P.~R.}\ \bibnamefont {Hemmer}}, \bibinfo {author} {\bibfnamefont
  {J.}~\bibnamefont {Wrachtrup}}, \ and\ \bibinfo {author} {\bibfnamefont
  {F.}~\bibnamefont {Jelezko}},\ }\href {\doibase 10.1126/science.1189075}
  {\bibfield  {journal} {\bibinfo  {journal} {Science}\ }\textbf {\bibinfo
  {volume} {329}},\ \bibinfo {pages} {542} (\bibinfo {year}
  {2010})}\BibitemShut {NoStop}%
\bibitem [{\citenamefont {Pfaff}\ \emph {et~al.}(2013)\citenamefont {Pfaff},
  \citenamefont {Taminiau}, \citenamefont {Robledo}, \citenamefont {Bernien},
  \citenamefont {Markham}, \citenamefont {Twitchen},\ and\ \citenamefont
  {Hanson}}]{pfaff_demonstration_2013}%
  \BibitemOpen
  \bibfield  {author} {\bibinfo {author} {\bibfnamefont {W.}~\bibnamefont
  {Pfaff}}, \bibinfo {author} {\bibfnamefont {T.~H.}\ \bibnamefont {Taminiau}},
  \bibinfo {author} {\bibfnamefont {L.}~\bibnamefont {Robledo}}, \bibinfo
  {author} {\bibfnamefont {H.}~\bibnamefont {Bernien}}, \bibinfo {author}
  {\bibfnamefont {M.}~\bibnamefont {Markham}}, \bibinfo {author} {\bibfnamefont
  {D.~J.}\ \bibnamefont {Twitchen}}, \ and\ \bibinfo {author} {\bibfnamefont
  {R.}~\bibnamefont {Hanson}},\ }\href {\doibase 10.1038/nphys2444} {\bibfield
  {journal} {\bibinfo  {journal} {Nat Phys}\ }\textbf {\bibinfo {volume} {9}},\
  \bibinfo {pages} {29} (\bibinfo {year} {2013})}\BibitemShut {NoStop}%
\bibitem [{\citenamefont {Blok}\ \emph {et~al.}(2015)\citenamefont {Blok},
  \citenamefont {Kalb}, \citenamefont {Reiserer}, \citenamefont {Taminiau},\
  and\ \citenamefont {Hanson}}]{blok_quantum_2015}%
  \BibitemOpen
  \bibfield  {author} {\bibinfo {author} {\bibfnamefont {M.~S.}\ \bibnamefont
  {Blok}}, \bibinfo {author} {\bibfnamefont {N.}~\bibnamefont {Kalb}}, \bibinfo
  {author} {\bibfnamefont {A.}~\bibnamefont {Reiserer}}, \bibinfo {author}
  {\bibfnamefont {T.~H.}\ \bibnamefont {Taminiau}}, \ and\ \bibinfo {author}
  {\bibfnamefont {R.}~\bibnamefont {Hanson}},\ }\href {\doibase
  10.1039/C5FD00113G} {\bibfield  {journal} {\bibinfo  {journal} {Faraday
  Discuss.}\ }\textbf {\bibinfo {volume} {184}},\ \bibinfo {pages} {173}
  (\bibinfo {year} {2015})}\BibitemShut {NoStop}%
\bibitem [{\citenamefont {Massar}\ and\ \citenamefont
  {Popescu}(1995)}]{massar_optimal_1995}%
  \BibitemOpen
  \bibfield  {author} {\bibinfo {author} {\bibfnamefont {S.}~\bibnamefont
  {Massar}}\ and\ \bibinfo {author} {\bibfnamefont {S.}~\bibnamefont
  {Popescu}},\ }\href {\doibase 10.1103/PhysRevLett.74.1259} {\bibfield
  {journal} {\bibinfo  {journal} {Phys. Rev. Lett.}\ }\textbf {\bibinfo
  {volume} {74}},\ \bibinfo {pages} {1259} (\bibinfo {year}
  {1995})}\BibitemShut {NoStop}%
\bibitem [{\citenamefont {Terhal}(2015)}]{terhal_quantum_2015}%
  \BibitemOpen
  \bibfield  {author} {\bibinfo {author} {\bibfnamefont {B.~M.}\ \bibnamefont
  {Terhal}},\ }\href {\doibase 10.1103/RevModPhys.87.307} {\bibfield  {journal}
  {\bibinfo  {journal} {Rev. Mod. Phys.}\ }\textbf {\bibinfo {volume} {87}},\
  \bibinfo {pages} {307} (\bibinfo {year} {2015})}\BibitemShut {NoStop}%
\bibitem [{\citenamefont {Barreiro}\ \emph {et~al.}(2011)\citenamefont
  {Barreiro}, \citenamefont {M{\"u}ller}, \citenamefont {Schindler},
  \citenamefont {Nigg}, \citenamefont {Monz}, \citenamefont {Chwalla},
  \citenamefont {Hennrich}, \citenamefont {Roos}, \citenamefont {Zoller},\ and\
  \citenamefont {Blatt}}]{barreiro_opensystem_2011}%
  \BibitemOpen
  \bibfield  {author} {\bibinfo {author} {\bibfnamefont {J.~T.}\ \bibnamefont
  {Barreiro}}, \bibinfo {author} {\bibfnamefont {M.}~\bibnamefont
  {M{\"u}ller}}, \bibinfo {author} {\bibfnamefont {P.}~\bibnamefont
  {Schindler}}, \bibinfo {author} {\bibfnamefont {D.}~\bibnamefont {Nigg}},
  \bibinfo {author} {\bibfnamefont {T.}~\bibnamefont {Monz}}, \bibinfo {author}
  {\bibfnamefont {M.}~\bibnamefont {Chwalla}}, \bibinfo {author} {\bibfnamefont
  {M.}~\bibnamefont {Hennrich}}, \bibinfo {author} {\bibfnamefont {C.~F.}\
  \bibnamefont {Roos}}, \bibinfo {author} {\bibfnamefont {P.}~\bibnamefont
  {Zoller}}, \ and\ \bibinfo {author} {\bibfnamefont {R.}~\bibnamefont
  {Blatt}},\ }\href {\doibase 10.1038/nature09801} {\bibfield  {journal}
  {\bibinfo  {journal} {Nature}\ }\textbf {\bibinfo {volume} {470}},\ \bibinfo
  {pages} {486} (\bibinfo {year} {2011})}\BibitemShut {NoStop}%
\bibitem [{\citenamefont {Pfaff}\ \emph {et~al.}(2014)\citenamefont {Pfaff},
  \citenamefont {Hensen}, \citenamefont {Bernien}, \citenamefont {{van Dam}},
  \citenamefont {Blok}, \citenamefont {Taminiau}, \citenamefont {Tiggelman},
  \citenamefont {Schouten}, \citenamefont {Markham}, \citenamefont {Twitchen},\
  and\ \citenamefont {Hanson}}]{pfaff_unconditional_2014}%
  \BibitemOpen
  \bibfield  {author} {\bibinfo {author} {\bibfnamefont {W.}~\bibnamefont
  {Pfaff}}, \bibinfo {author} {\bibfnamefont {B.~J.}\ \bibnamefont {Hensen}},
  \bibinfo {author} {\bibfnamefont {H.}~\bibnamefont {Bernien}}, \bibinfo
  {author} {\bibfnamefont {S.~B.}\ \bibnamefont {{van Dam}}}, \bibinfo {author}
  {\bibfnamefont {M.~S.}\ \bibnamefont {Blok}}, \bibinfo {author}
  {\bibfnamefont {T.~H.}\ \bibnamefont {Taminiau}}, \bibinfo {author}
  {\bibfnamefont {M.~J.}\ \bibnamefont {Tiggelman}}, \bibinfo {author}
  {\bibfnamefont {R.~N.}\ \bibnamefont {Schouten}}, \bibinfo {author}
  {\bibfnamefont {M.}~\bibnamefont {Markham}}, \bibinfo {author} {\bibfnamefont
  {D.~J.}\ \bibnamefont {Twitchen}}, \ and\ \bibinfo {author} {\bibfnamefont
  {R.}~\bibnamefont {Hanson}},\ }\href {\doibase 10.1126/science.1253512}
  {\bibfield  {journal} {\bibinfo  {journal} {Science}\ }\textbf {\bibinfo
  {volume} {345}},\ \bibinfo {pages} {532} (\bibinfo {year}
  {2014})}\BibitemShut {NoStop}%
\end{thebibliography}%

\section*{Acknowledgements}
The authors thank M. Bakker for experimental assistance and P. C. Humphreys, V. V. Dobrovitski and S. B. van Dam for critically reading the manuscript.

\section*{Contributions}
NK and THT devised the experiment. NK, JC and THT prepared and characterized the experimental apparatus. DJT and MM grew the diamond substrate. NK collected and analysed the data with the help of RH and THT. NK and THT wrote the manuscript with input from all authors.

\clearpage

\begin{figure}[h!]
	\includegraphics[width = 150 mm]{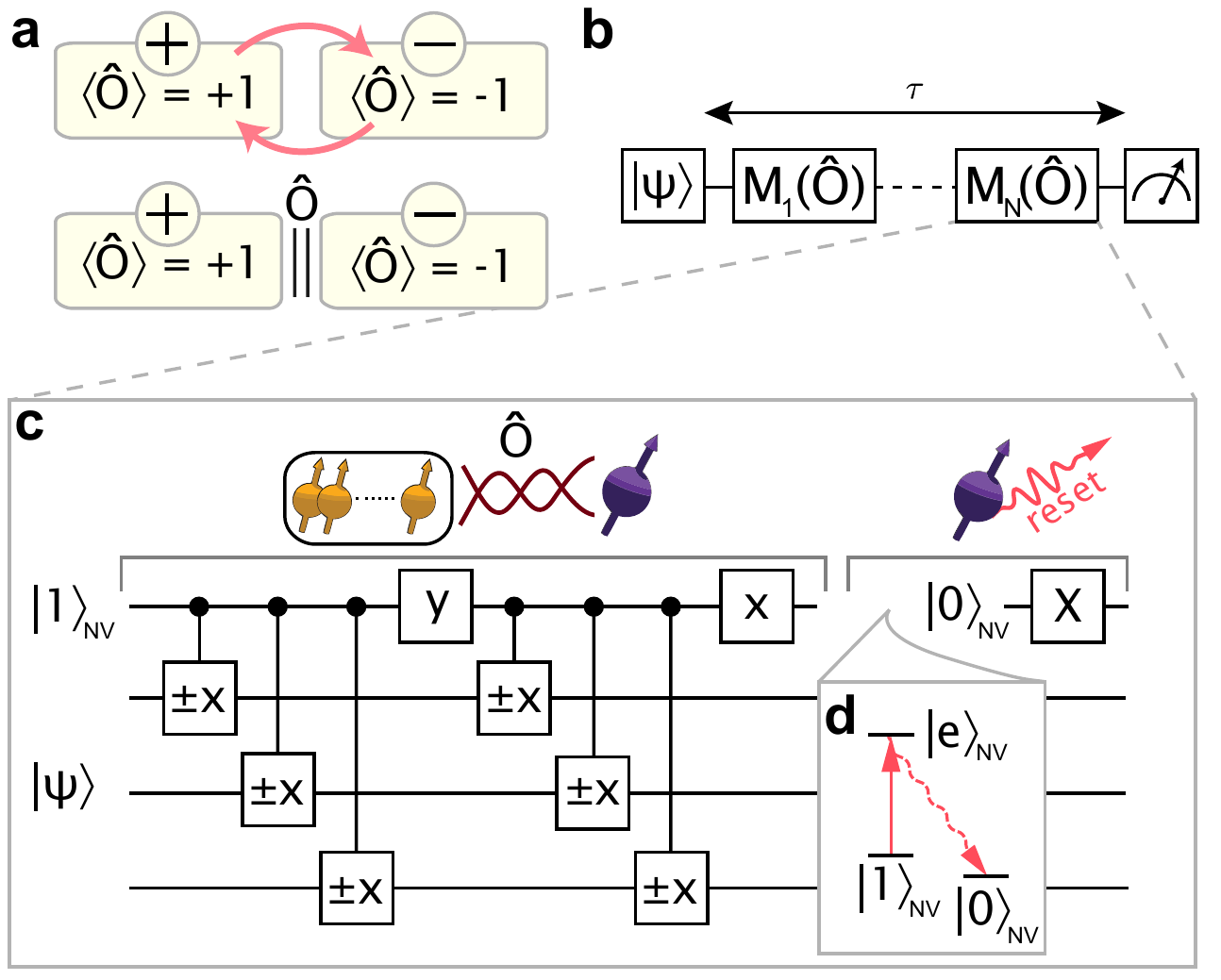}
	\caption{\label{fig:1} \textbf{Concept and experimental sequence. (a)} Quantum Zeno subspaces. The state space of a quantum system is divided into two subspaces (yellow boxes) of an observable $\hat{O}$. Coherent transitions between the two subspaces occur while the system is unperturbed (top, red arrows) but are strongly inhibited if $\hat{O}$ is repeatedly projected (bottom). \textbf{(b)} Experimental sequence. After initialization in $\ket{\psi}$, $N$ equidistantly distributed projections $M(\hat{O})$ (see Eq. (1)) are applied during a total evolution time $\tau$ and the state of the system is read out. \textbf{(c)} Realization of $M(\hat{O} = \hat{\sigma}_\mathrm{x} \hat{\sigma}_\mathrm{x} \hat{\sigma}_\mathrm{x})$ for three nuclear spins. First the state of the nuclear spins (yellow) is entangled with the ancilla electron-spin state (purple). Second the electron spin is projected and reinitialized in $\ket{1}_\mathrm{NV}$ (see also panel d) through optical pumping ($30\,\mathrm{\mu s}$) to  $\ket{0}_\mathrm{NV}$ and a subsequent microwave $\pi$-pulse (X). The x and y gates are $\pi /2$ rotations around the X- and Y-axis respectively. Controlled gates indicate that the direction is determined by the electron spin [27]. See Supplementary Fig. 2 for pulse sequences for projections on one and two spins. \textbf{(d)} Relevant electron spin levels for optical repumping through selective resonant excitation of $\ket{1}_\mathrm{NV}$ to $\ket{e}_\mathrm{NV}$. We prepare the nuclear spin states in the $\langle \hat{O} \rangle = +1$ subspace and associate this subspace to the electron state $\ket{0}_\mathrm{NV}$ in the entangling sequence so that the optical projection ideally never excites the NV center.} 
\end{figure}
\clearpage

\begin{figure}[h!]
	\includegraphics[width = 150 mm]{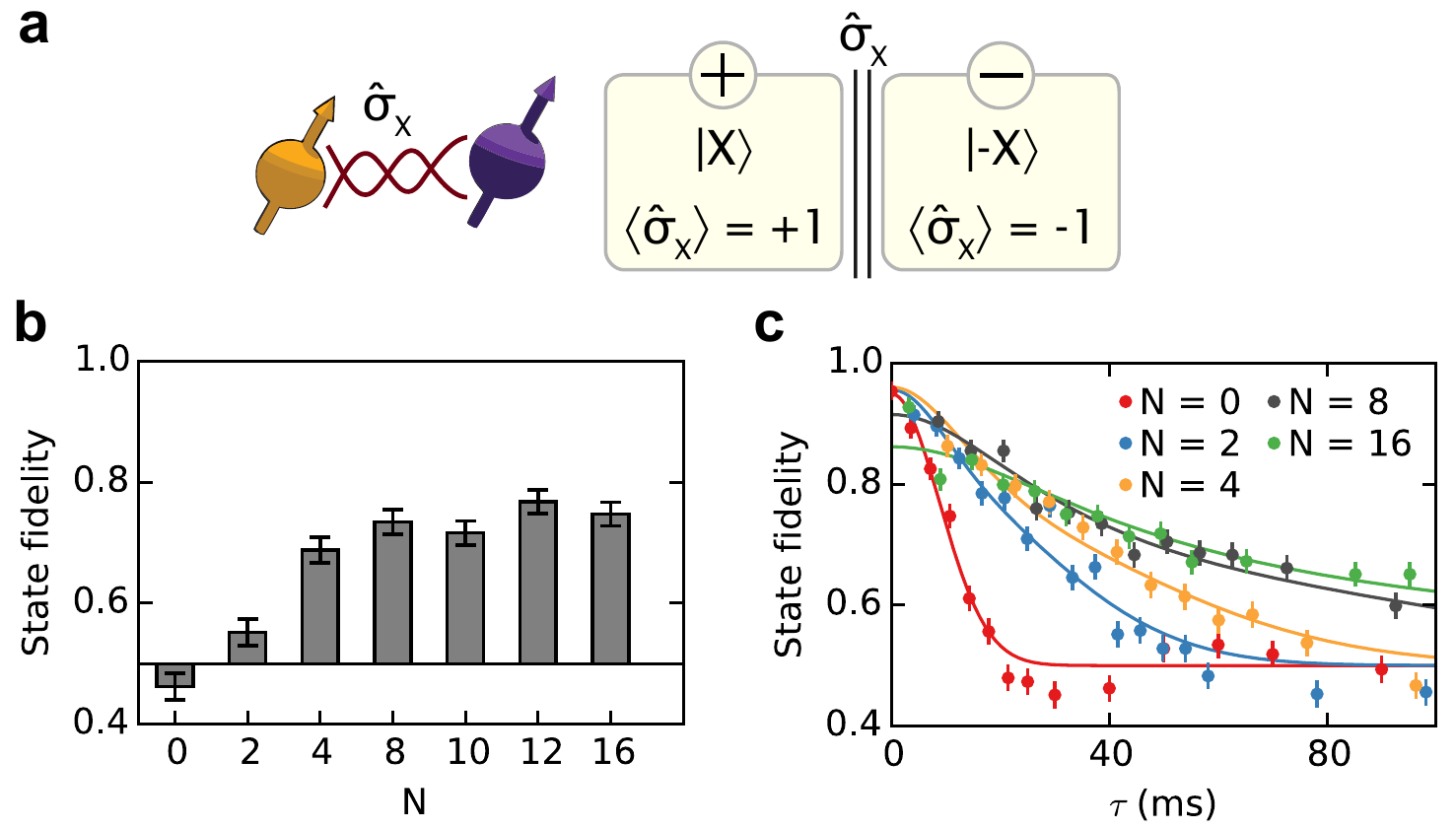}
	\caption{\label{fig:2} \textbf{Quantum Zeno effect for a single spin superposition state. (a)} Quantum Zeno subspaces for a single nuclear spin (spin 1) and $\hat{O} = \hat{\sigma}_\mathrm{x}$. Each eigenspace of $\hat{\sigma}_\mathrm{x}$ consists of one state ($\ket{X}$ or $\ket{-X}$) with the respective eigenvalue indicated by the circled $+/-$ signs. \textbf{(b)} State fidelity for $\ket{X}$ after $\tau = 40\,\mathrm{ms}$. The fidelity increases with the number of projections $N$ until $N \approx 16$ projections. \textbf{(c)} The complete time traces for the storage of $\ket{X}$ show that the dephasing time increases with the number of projections. The curves are fits to the theoretically expected fidelity (see Eq. (3)). All data is corrected for the read-out fidelity (Supplementary Fig. 3 and Supplementary Note 3). All error bars are $1$ s.d.} 
\end{figure}
\clearpage

\begin{figure}[h!]
	\includegraphics[width = 150 mm]{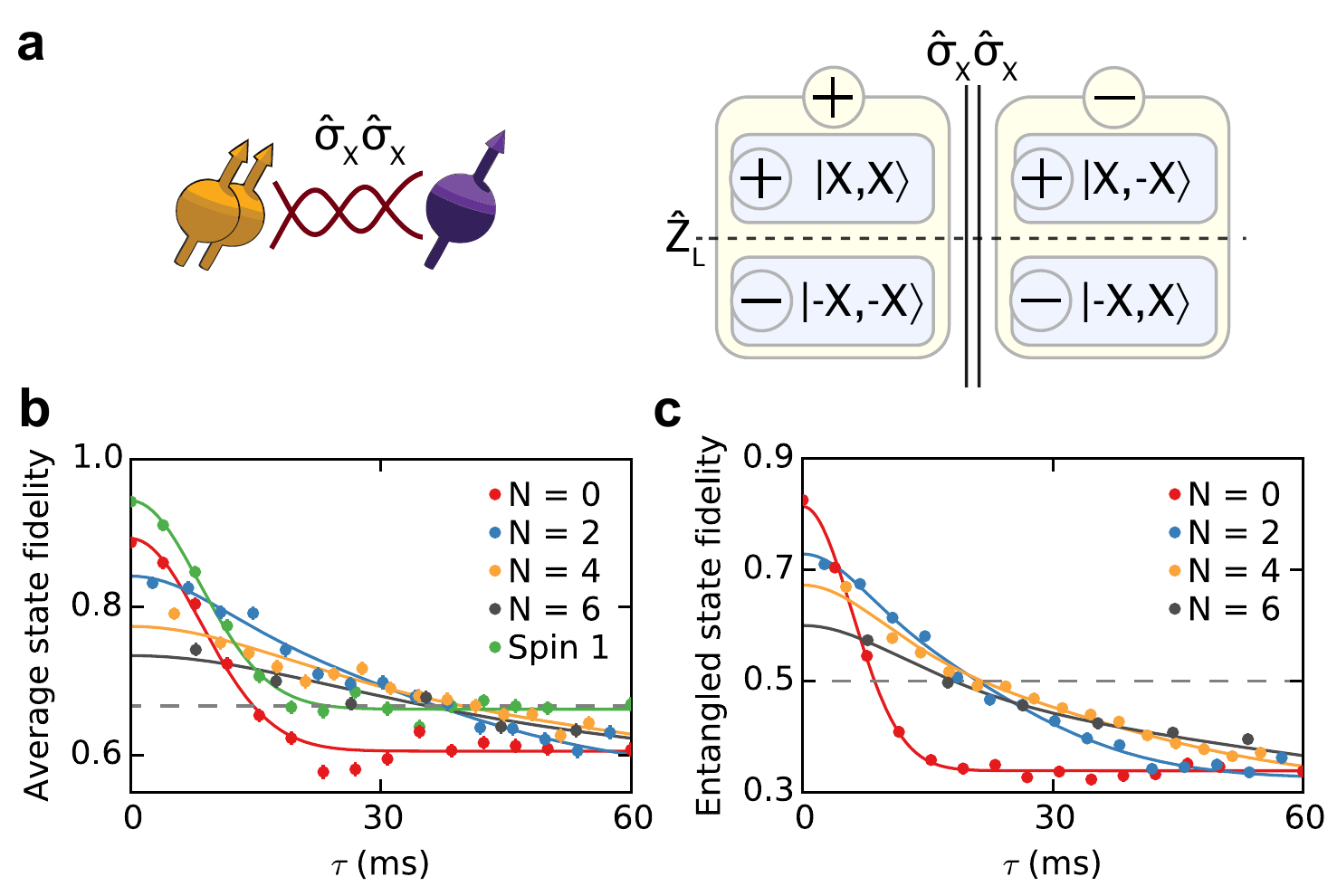}
	\caption{\label{fig:3} \textbf{Storing a logical quantum bit by repeated two-spin projections. (a)}  Schematic: the four-dimensional state-space of two $^\mathrm{13}$C spins (spin 1 and 2) is divided into two subspaces by repetitively projecting $\hat{O} = \hat{\sigma}_\mathrm{x} \hat{\sigma}_\mathrm{x}$ through entanglement with the ancilla spin. We define a logical quantum bit with logical operator $\hat{Z}_\mathrm{L} = \hat{\sigma}_\mathrm{x} \hat{I}$ (dashed line). \textbf{(b)} Storing a logical quantum bit. The average logical state fidelity for the six logical input states as a function of time and for a varying number of projections $N$, compared to the nuclear spin with the longest dephasing time (spin 1).  To compare the results to the best possible decay for the single nuclear spin, we eliminate potential systematic detunings by measuring $\sqrt{\langle \hat{\sigma}_\mathrm{x} \rangle ^2 + \langle \hat{\sigma}_\mathrm{y} \rangle ^2}$. The dashed horizontal line is the threshold of $2/3$ for classically storing quantum states [34]. \textbf{(c)} Preserving two-spin entangled states. The two-spin state fidelity, averaged over the four entangled input states, indicates that general two-spin states in the subspace are preserved. Above the dashed horizontal line ($F = 0.5$) the state is entangled. For $N= 2$, $4$ and $6$ projections, entanglement is preserved longer than without projections. Solid lines are fits to Eq. (3) with the initial amplitude $A$, the offset and the effective dephasing time $T_{2,\mathrm{eff}}^*$ as free parameters. Error bars are $1$ s.d. and are smaller than the symbols.} 
\end{figure}

\clearpage
\begin{figure}[h!]
	\includegraphics[width = 85 mm]{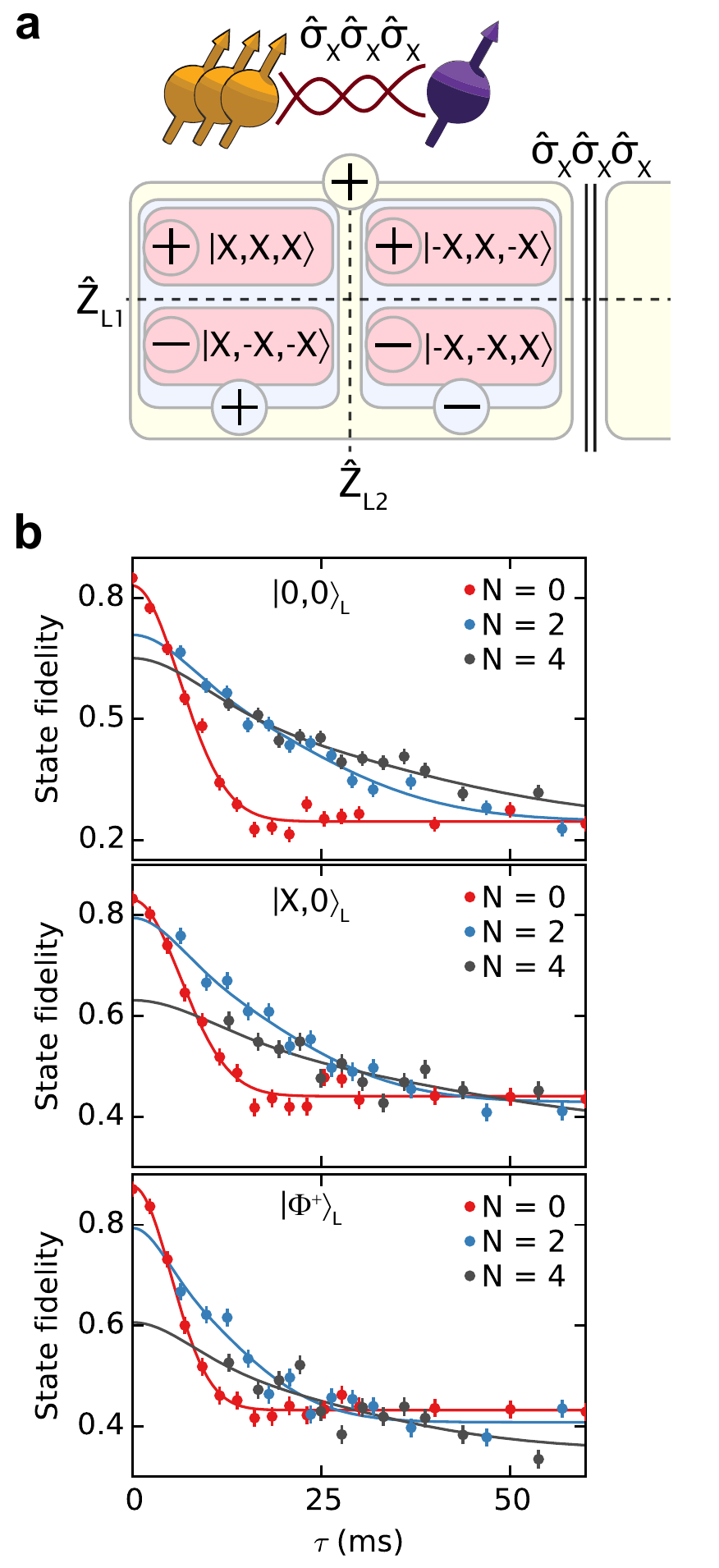}
	\caption{\label{fig:4} \textbf{Two logical qubits in a quantum Zeno subspace. (a)}  Schematic for three nuclear spins (spins 1,2 and 3) and $\hat{O} = \hat{\sigma}_\mathrm{x} \hat{\sigma}_\mathrm{x} \hat{\sigma}_\mathrm{x}$. Two four-dimensional subspaces are created (yellow box). For simplicity we only show the positive subspace, which contains states of the form $\alpha \ket{0,0}_\mathrm{L} + \beta \ket{0,1}_\mathrm{L}+ \gamma \ket{1,0}_\mathrm{L}+ \delta \ket{1,1}_\mathrm{L}$. Within this subspace two logical qubits are defined by the logical operators $\hat{Z}_\mathrm{L1} = \hat{\sigma}_\mathrm{x} \hat{I} \hat{\sigma}_\mathrm{x}$, $\hat{X}_\mathrm{L1} = \hat{I} \hat{\sigma}_\mathrm{z} \hat{\sigma}_\mathrm{z}$ and $\hat{Z}_\mathrm{L2} = \hat{I} \hat{\sigma}_\mathrm{x} \hat{\sigma}_\mathrm{x}$, $\hat{X}_\mathrm{L2} = \hat{\sigma}_\mathrm{z} \hat{I} \hat{\sigma}_\mathrm{z}$ (blue and red boxes). Plus and minus signs indicate eigenvalues of the associated operator. \textbf{(b)} Logical state fidelities for three logical states: eigenstate $\ket{0,0}_\mathrm{L}$, superposition state $\ket{X,0}_\mathrm{L}$, and the entangled state $\ket{\Phi^+}_\mathrm{L}$. The results show that repeated projections of the three-spin operator $\hat{O} = \hat{\sigma}_\mathrm{x} \hat{\sigma}_\mathrm{x} \hat{\sigma}_\mathrm{x}$ preserve the two logical qubits while inhibiting dephasing. Solid lines are fits to Eq. (3). The fidelities decay to different values for large $\tau$ because $\ket{\Phi^+}_\mathrm{L}$ and $\ket{X,0}_\mathrm{L}$ are eigenstates of operators of the form $\hat{I} \hat{\sigma}_\mathrm{z} \hat{\sigma}_\mathrm{z}$ or one of its permutations, whose expectation values are unaffected by dephasing. Error bars are $1$ s.d.} 
\end{figure}

\clearpage
\begin{figure}[h!]
	\includegraphics[width = 85 mm]{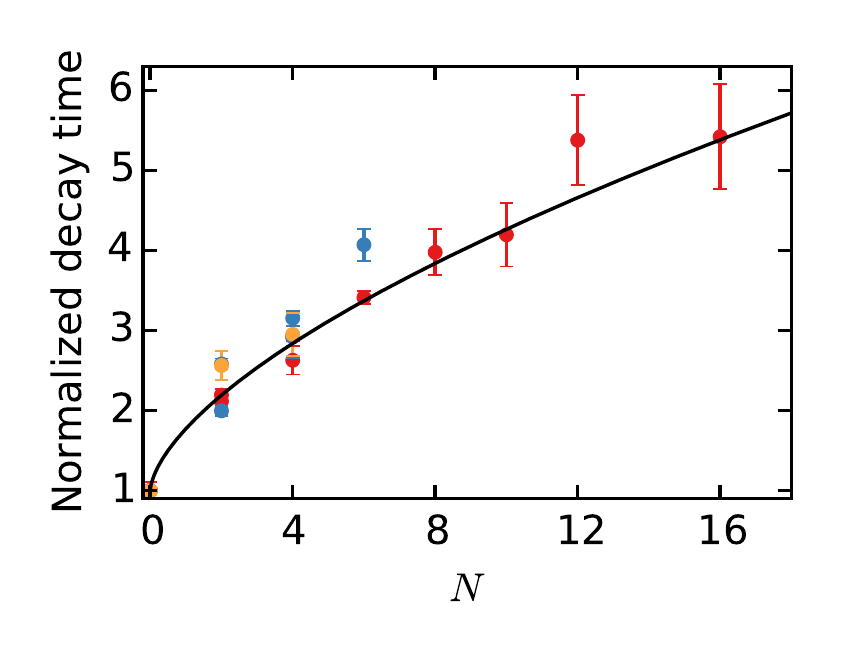}
	\caption{\label{fig:5} \textbf{Scaling of the decay time with increasing number of projections.}  The fitted decay times for all measurements in this article (Figs. 2-4) are compared to the theoretical decay time enhancement (solid line). All values are taken relative to the value without projections ($N=0$). The data are averaged according to the number of operators in the expectation values that are subject to dephasing (i.e. the number of $\hat{\sigma}_\mathrm{x}$ and/or $\hat{\sigma}_\mathrm{y}$). Red: One operator. Blue: Two operators. Orange: Three operators (see Supplementary Fig. 4 for raw data). To show that the normalized decay time is independent of the number of nuclear spins, we distinguish data with a differing total number of nuclear spins. For instance, measurements of $\langle \hat{\sigma}_\mathrm{x} \rangle$(with $\hat{O} = \hat{\sigma}_\mathrm{x}$) or $\langle \hat{\sigma}_\mathrm{x} \hat{I} \rangle$ (with $\hat{O} = \hat{\sigma}_\mathrm{x} \hat{\sigma}_\mathrm{x}$) are represented by separate data points. The theory curve is obtained by evaluating Eq. (3) up to $N=16$ (see Supplementary Fig. 5). Error bars are 1 s.d.} 
\end{figure}
\clearpage
\section*{Supplementary Information}

\beginsupplement
\begin{figure}[h!]
	\includegraphics[width = 150 mm]{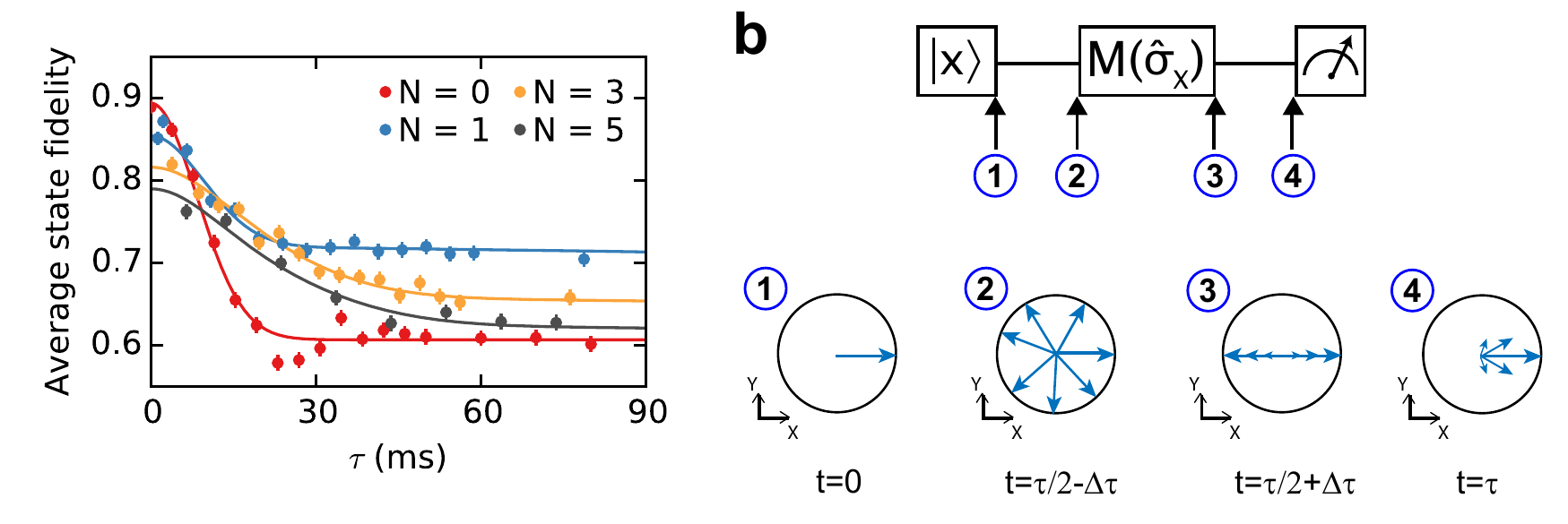}
	\caption{\label{fig:OddMeasurements} Analysis for an odd number of projections. (a) Experimental results for the logical qubit in the two-spin case (Fig. 3b of the main text). We observe a clear elevation of the average state fidelity for times much longer than $T_2^*$ for 1 (blue) and 3 (orange) projections. This was also observed in J. Cramer et al. for $N = 1$ \cite{cramer_repeated_2015}. The behavior for times much longer than the decoherence time is fully explained by our model and can be accurately fit with \eref{finalcountdown}. (b) Schematic explanation of the elevated signal for large evolution times $\tau \gg T_2^* $. Top panel: A spin-$1/2$ particle is initialized in a superposition state $\ket{X}$, evolves freely for a time $\tau/2$ until it is projected ($M\left(\hat{\sigma}_\mathrm{x} \right)$) onto $\ket{\pm X}$ and is finally read out after a total evolution time of $\tau$. Bottom panel: Sketch of an ensemble of states in the XY plane of the Bloch sphere. The state of the particle is well defined after initialization \textcircled{1}. The state is however completely mixed due to a random frequency detuning before projection \textcircled{2} if the free evolution $\tau/2$ is much longer than the dephasing time $T_2^*$. Projection into $\ket{\pm X}$ effectively projects the Bloch vector onto the x-axis \textcircled{3}. The frequency detuning of the particle remains constant after projection such that the ensemble average partly rephases (similarly to a spin echo) after another evolution time of $\tau/2$. We call this effect a filter since a single projection of $\hat{\sigma}_\mathrm{x}$ completely mixes the state if the particle is in $\ket{ \pm Y} = (\ket{X} \pm i\ket{-X})/\sqrt{2}$, effectively filtering those cases out. In contrast states which are affected by a detuning but got rotated to $\ket{\pm X}$ at the time of the projection are unaffected.
	} 
\end{figure}

\begin{figure}[h!]
	\includegraphics{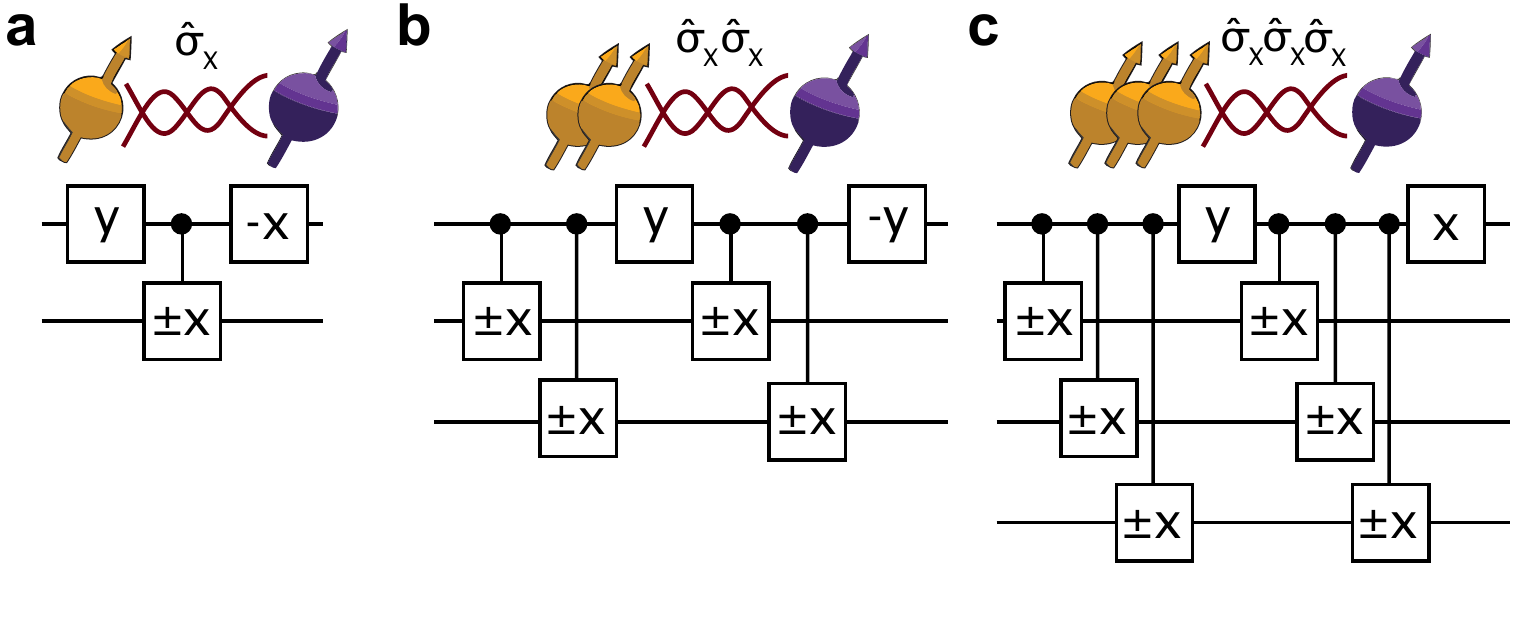}
	\caption{\label{fig:Gate_sequence} Gate sequences to project the observables $\hat{\sigma}_\mathrm{x}$ (a), $\hat{\sigma}_\mathrm{x}\hat{\sigma}_\mathrm{x}$ (b) and $\hat{\sigma}_\mathrm{x}\hat{\sigma}_\mathrm{x}\hat{\sigma}_\mathrm{x}$ (c). $\mathrm{x}$ and $\mathrm{y}$ are $\pi/2$ rotations around the X and Y axis with the orientation given by the sign. We use a combination of dynamical decoupling sequences that act as electron-state-conditional quantum gates on the $\mathrm{^{13}C}$ spins \cite{taminiau_universal_2014} to address them (controlled gates in all panels).}
\end{figure}

\begin{figure}[h!]
	\includegraphics[width=89mm]{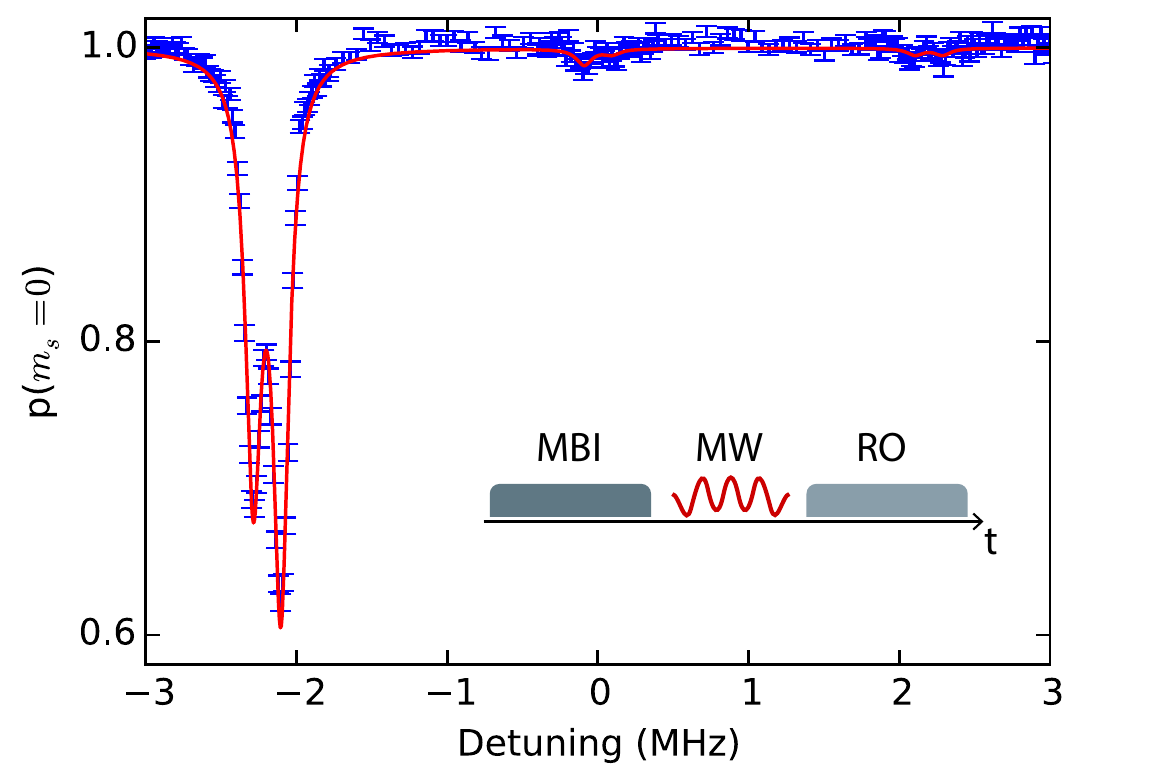}
	\caption{\label{fig:Nitrogen_population} Electron spin resonance to determine the initialization fidelity of the $\mathrm{^{14}N}$ nuclear spin ($I = 1$) in $m_I = -1$. Inset: Schematic of the experimental sequence. After measurement-based initialization of the nucleus (MBI) a weak microwave pulse (MW) \cite{pfaff_unconditional_2014} with a duration of $8\,\mathrm{\mu s}$ is applied and followed by optical read out (RO) of the electron spin. The detuning is relative to the central microwave frequency ($1.74667(1)\,\mathrm{GHz}$) of the $m_s = 0$ to $m_s = -1$ transition of the electron spin. The transition is split into six lines due to strong hyperfine coupling to the $\mathrm{^{14}N}$ nuclear spin (with a coupling strength of $2\pi \cdot 2.195(1)\,\mathrm{MHz}$) and a $\mathrm{^{13}C}$ nuclear spin ($I = 1/2$, with a coupling strength of $2\pi \cdot 182(1)\,\mathrm{kHz}$) in the vicinity of the NV centre. The data is therefore fit to six Lorentzian functions with variable width, spacing and amplitude. From the fitted amplitudes we extract the population of the $\mathrm{^{14}N}$ nuclear spin after MBI: $p_{-1} = 0.96(1) \quad p_{0} = 0.022(8) \quad p_{+1} = 0.014(8)$.}
\end{figure}

\begin{figure}[h!]
	\includegraphics[width = 150mm]{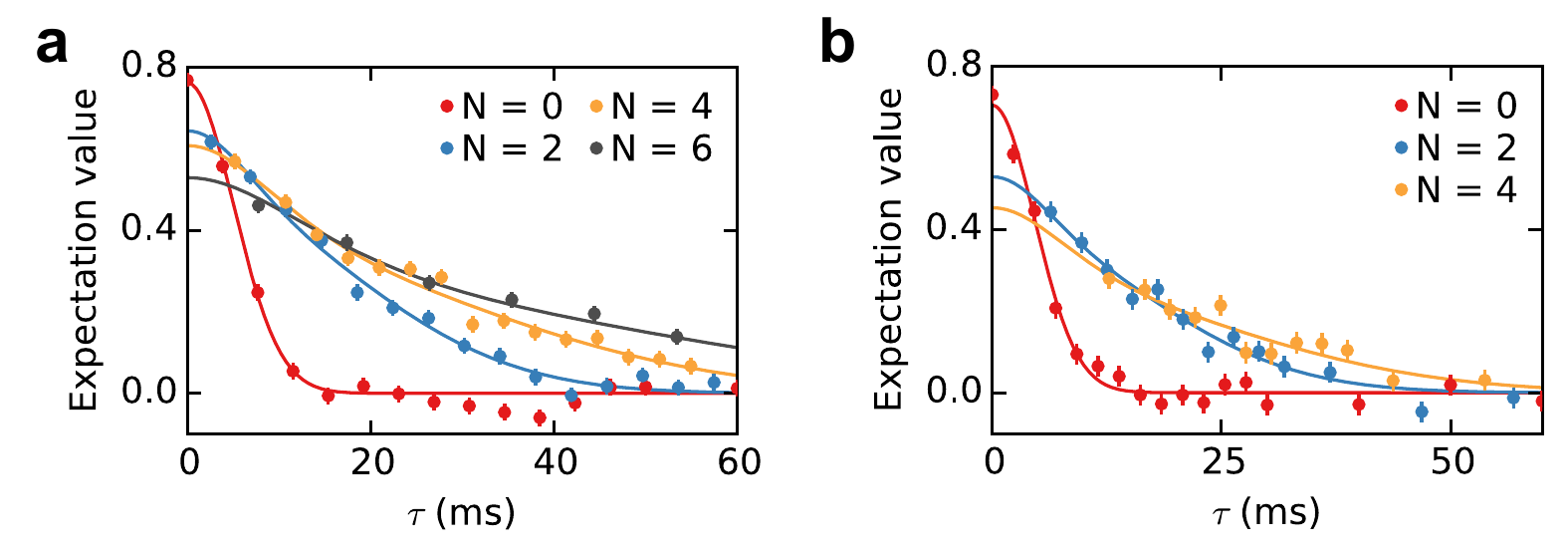}
	\caption{\label{fig:XXX_Contrast}(a) Decay of $\langle \hat{\sigma}_\mathrm{x} \hat{\sigma}_\mathrm{x} \rangle$ for a varying number of projections. The data are averaged over six input states that correspond to the states used for Fig. 3b of the main text. The number of projections N is given in the legend. (b) Decay of $\langle \hat{\sigma}_\mathrm{x} \hat{\sigma}_\mathrm{x} \hat{\sigma}_\mathrm{x} \rangle$. The data are averaged over the three states of Fig. 4b of the main text. The fitted $1/\sqrt{e}$-time for $N=0$ is $4.6(2)\,\mathrm{ms}$. The amount of projections for each data trace is given in the legend. The fitted, relative, decay constants for these data are included in Fig. 5 of the main text.}
\end{figure}

\begin{figure}[h!]
	\centering
	\includegraphics[width=89mm]{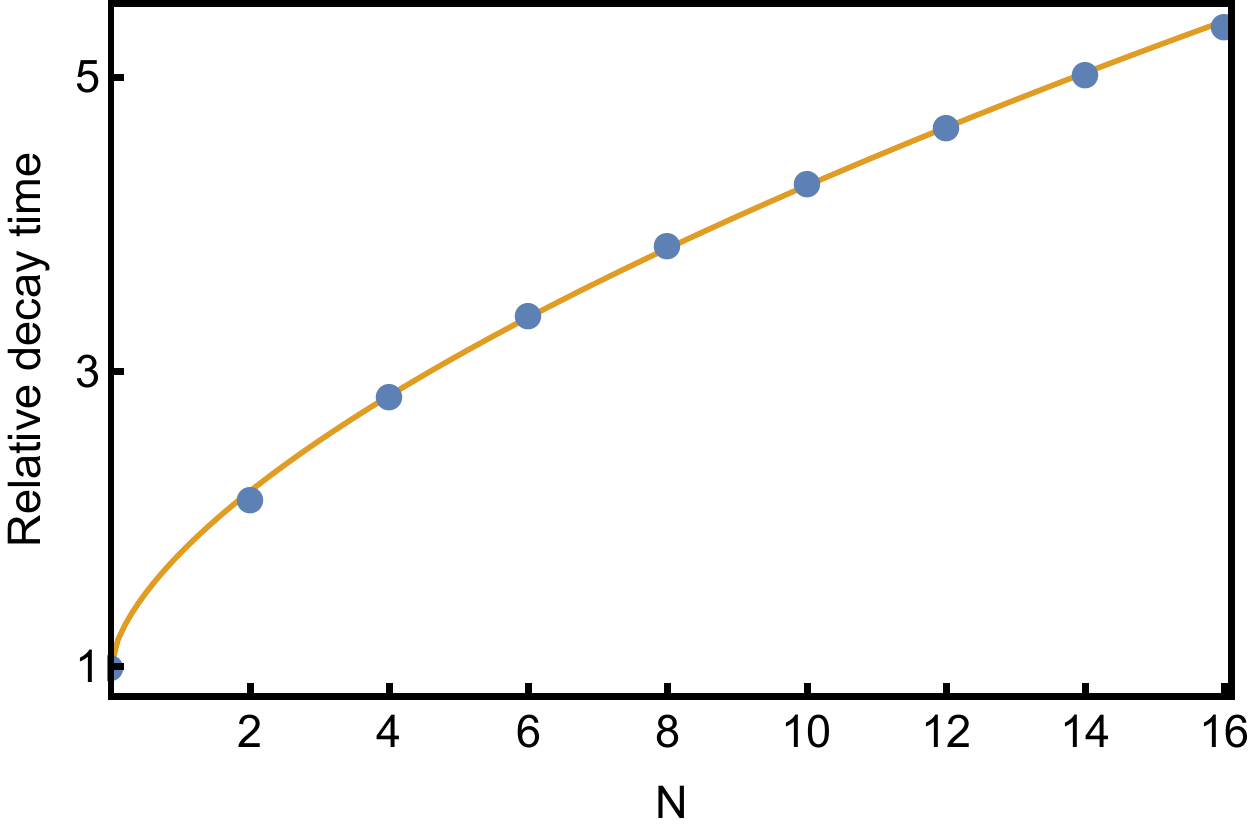}
	\caption{Blue data: extracted $\frac{1}{\sqrt{e}}$-time from the analytical solutions to \eref{finalcountdown}. Orange line: Fitting the extracted times to the power law $1+\mu M^\nu$ yields $\mu = 0.77(1)$ and $\nu = 0.63(1)$ and gives a good approximation for the experimentally investigated number of projections $N$.}
	\label{fig:DecayTheory}
\end{figure}

\clearpage

\begin{table}[h!]
	\centering
	
	\begin{tabular}{c|c|c}
		& fitted $T_2^*$ (ms) & expected $T_{2,eff}^*$ (ms) \\
		\hline
		$C_1$         & $12.4(9)$              & /         \\
		$C_2$         & $8.2(7)$             & /         \\
		$C_3$         & $21(1)$               & /         \\
		$C_1C_2$      & $7.1(3)$              & $6.8(3)$  \\
		$C_1C_2C_3$   & $6.5(2)$              & $6.5(4)$                              
	\end{tabular}
	\caption{Comparison of fitted and expected decay time $T_2^*$ for multi-qubit observables. The expected $T_2^*$ value is calculated by assuming uncorrelated noise and using $\left(1/T_{2,\mathrm{eff}}^*\right)^2 = \sum_i\left(1/T_{2,i}^*\right)^2$ (see Supplementary Note 3). We find agreement between expected and measured decay times.}
	\label{tbl:FitValues}
\end{table}

\section*{Supplementary Note 1: Device characteristics}
The experiments were conducted in a confocal microscope at cryogenic temperatures ($4\,\mathrm{K}$). The investigated sample is a chemical-vapour-deposition homoepitaxially grown diamond of type IIa with a natural composition of carbon isotopes. The diamond has been cut along the $<111>$ crystal axis and was grown by Element Six.

\section*{Supplementary Note 2: Analytical model and data processing}
\subsubsection{Derivation of the analytical model}
This section outlines the derivation of Eq. (3) of the main text. We consider $k$ two-level systems with random, uncorrelated and constant detuning $\Delta_i$ for each two-level system. Without loss of generality, the initial ($t=0$) state is chosen to be the balanced superposition state $\ket{X_1 \cdots X_k}$. This state is an eigenstate of the projected operator $\hat{\sigma}_\mathrm{x,1} \cdots \hat{\sigma}_\mathrm{x,k}$. After the first projection at time $t$ and evolution of the system for another time $t$ an analytic expression for the expectation value $\langle \hat{\sigma}_\mathrm{x,1} \cdots \hat{\sigma}_\mathrm{x,k} \rangle_{N=1} = \mathrm{Tr}\left[\rho \hat{\sigma}_\mathrm{x,1} \cdots \hat{\sigma}_\mathrm{x,k}  \right]$, with the density matrix $\rho$, can be derived.

\be
\langle \hat{\sigma}_\mathrm{x,1} \cdots \hat{\sigma}_\mathrm{x,k} \rangle_{N=1} = \frac{1}{2^{k}} \sum_{\alpha_1 \in \{-1 , +1 \} } \cdots \sum_{\alpha_N \in \{ -1 , +1\} } \cos^2[(\sum\limits_{i=1}^{k}\alpha_i \Delta_i)t] = \frac{1}{2^{k-1}} \sum_{\alpha_2 ... \alpha_k} \cos^2[(\Delta_1 + \sum\limits_{i=2}^{k}\alpha_i \Delta_i)t] \, .
\label{eq:Onerun}
\ee

With the sum over all possible combinations of relative detunings by choosing the binary values of $\alpha_i = \pm 1 \,\, \forall \,\, i \in \{1,2,...,k\}$. Note that the notation for the sum over all configurations of $\alpha_i$ has been simplified for the last equality. This sum of cosine terms originates from static terms in the relevant entries of the density matrix after projection. The formula for a single projection ($N=1$) is then readily extended to $N$ projections
\be
\langle \hat{\sigma}_\mathrm{x,1} \cdots \hat{\sigma}_\mathrm{x,k} \rangle_N = \frac{1}{2^{k-1}} \sum_{\alpha_2 ... \alpha_k} \cos^{N+1}[(\Delta_1 + \sum\limits_{i=2}^{k}\alpha_i \Delta_i)t].
\ee

We obtain the ensemble average $\overline{\langle \hat{\sigma}_{x,1} \cdots \hat{\sigma}_{x,k} \rangle}_N$ by integration over a normal distribution $G_i[\Delta_i]$ of width $\sigma_i =  \sqrt{2}/T^*_{2,i}$ for each $\Delta_i$
\be
\overline{\langle \hat{\sigma}_\mathrm{x,1} \cdots \hat{\sigma}_\mathrm{x,k} \rangle}_N = \int \cdots \int \langle \hat{\sigma}_\mathrm{x,1} \cdots \hat{\sigma}_\mathrm{x,k} \rangle_N \prod_{i = 1}^{k} G_i[\Delta_i] d\Delta_i.
\label{eq:Mean}
\ee

We single out one summand of $\langle \hat{\sigma}_\mathrm{x,1} \cdots \hat{\sigma}_\mathrm{x,k} \rangle_N$ with general $\alpha_i$ and perform the integration. The $\cos^{N+1}$ term is rewritten by using Euler's formula and the Binomial theorem

\be
\cos^{N+1}[(\Delta_1+\sum_{i=2}^{k}\alpha_i \Delta_i)t] = \frac{1}{2^{N+1}}\sum^{N+1}_{l=0}
\binom{N+1}{l}\exp\left[i (\Delta_1+\sum \alpha_i \Delta_i))t \right]^{N+1-2l}.
\label{eq:EulerBinom}
\ee

After inserting this rewritten cosine term into \eref{Mean} we obtain a product of Fourier transformations for each summand in \eref{EulerBinom}. It becomes clear that the precise assignment of the $\alpha_i$ coefficients does not play a role for the evaluation of the integral since the Fourier transformation and inverse Fourier transformation of a Gaussian give the same result: i.e. the precise assignment of $\alpha_i = \pm 1$ does not play a role for the evaluated integral. The normalization factor of $\frac{1}{2^{k-1}}$ therefore drops out when summing over all possible configurations of $\alpha_i$. Evaluating \eref{Mean} results in an analytic expression for the ensemble average $\overline{\langle \hat{\sigma}_\mathrm{x,1} \cdots \hat{\sigma}_\mathrm{x,k} \rangle}_N$
\be
\overline{\langle \hat{\sigma}_\mathrm{x,1} \cdots \hat{\sigma}_\mathrm{x,k} \rangle}_N = \frac{1}{2^{N+1}}\sum\limits_{l=0}^{N+1}\binom{N+1}{l}\exp \left[ -((N+1-2l)t)^2\sum_{i=1}^k (1/T_{2,i}^{*})^{2}\right] \, .
\label{eq:finalcountdown}
\ee

\eref{finalcountdown} describes the expected decay curve for a joint k-partite observable after $N$ joint projections. All operations are separated by the same duration $t$. Involving multiple nuclear spins results in an effective decay time $(1/T^*_{2,\mathrm{eff}})^2 = \sum_i (1/T_{2,i}^*)^2$ (as expected from the convolution of two Normal distributions). Correlations which are only partially subject to dephasing, e.g. $\langle \hat{Z}_1 \hat{Y}_2 \rangle$, only incorporate the relevant decoherence times (for the given example: $T^*_{2,\mathrm{eff}} = T^*_{2,2}$). In the main text, $t$ is replaced by the total evolution time $\tau =  (N+1)t$.

\subsubsection{Fitting routine}
The fits to the data are performed in the following way. First, the decay without projections ($N=0$) is fit with a Gaussian function. The initial amplitude, offset and width are extracted. Second, in order to fit data sets with multiple projections \eref{finalcountdown} is multiplied with the extracted amplitude and the offset is added (the offset originates from constant $\langle \hat{\sigma}_\mathrm{z,1} \cdots \hat{\sigma}_\mathrm{z,k} \rangle$ correlations that are not subject to dephasing and play a role when determining average state fidelities). The data set is then fitted with three free parameters: $T^*_{2,\mathrm{eff}}$, a global amplitude damping that parametrizes errors due to the added complexity of the experiment and the aforementioned constant offset.

\subsubsection{Scaling law}
In order to obtain the theory curve in Fig. 5 of the main text, we compute the theoretical enhancement of the dephasing time for a given number of projections. We calculate the normalized $\frac{1}{\sqrt{e}}$-time of \eref{finalcountdown} for $N$ being even and smaller than $17$. A modified scaling law, $1+\mu N^\nu$, is fit to the extracted characteristic dephasing times and the parameters $\mu = 0.77(1)$ and $\nu = 0.63(1)$ (see \fref{DecayTheory}) are found.


\section*{Supplementary Note 3: $\mathrm{^{13}C}$ spin read-out correction}
We correct the read-out results for errors introduced by the final conditional gates on the $\mathrm{^{13}C}$ spins to obtain the actual state fidelity. We employ a characterization technique developed in reference \cite{cramer_repeated_2015} and determine correction factors ($C_{C_i}$) for one-, two- and three-spin expectation values. The applied correction assumes a symmetric initialization and read-out process as well as a constant loss of fidelity due to imperfect initialization of the $\mathrm{^{14}N}$ spin of the NV centre. The probability to find the $\mathrm{^{14}N}$ spin in $m_I = -1$ after initialization is found to be 0.96(1) (see \fref{Nitrogen_population}). We obtain the following correction factors
\begin{align*}
C_{C_1} & = 0.94(1) & C_{C_2} & = 0.94(1) & C_{{C_1}{C_2}} & = 0.93(2) \\ 
C_{{C_1}{C_2}{C_3}} & = 0.90(2) & C_{{C_1}{C_3}} & = 0.93(2) & C_{{C_2}{C_3}} & = 0.95(2)\\
\end{align*}

\end{document}